\documentclass[sigconf]{acmart}
\usepackage{booktabs}
\usepackage{longtable}
\usepackage{multirow}
\usepackage{adjustbox}

\AtBeginDocument{%
  }

\begin{document}

\title{Benchmark for Evaluation and Analysis of Citation Recommendation Models}
\pagestyle{plain}


\author{Puja Maharjan}
\affiliation{%
  \institution{University of Konstanz}
  \city{Konstanz}
  \country{Germany}}
\email{puja.maharjan@uni-konstanz.de}

\renewcommand{\shortauthors}{Maharjan et al.}
\renewcommand\footnotetextcopyrightpermission[1]{}


\settopmatter{printacmref=false}
\begin{abstract}
Citation recommendation systems have attracted much academic interest, resulting in many studies and implementations. These systems help authors automatically generate proper citations by suggesting relevant references based on the text they have written. However, the methods used in citation recommendation differ across various studies and implementations. Some approaches focus on the overall content of papers, while others consider the context of the citation text. Additionally, the datasets used in these studies include different aspects of papers, such as metadata, citation context, or even the full text of the paper in various formats and structures. The diversity in models, datasets, and evaluation metrics makes it challenging to assess and compare citation recommendation methods effectively. To address this issue, a standardized dataset and evaluation metrics are needed to evaluate these models consistently. Therefore, we propose developing a benchmark specifically designed to analyze and compare citation recommendation models. This benchmark will evaluate the performance of models on different features of the citation context and provide a comprehensive evaluation of the models across all these tasks, presenting the results in a standardized way. By creating a benchmark with standardized evaluation metrics, researchers and practitioners in the field of citation recommendation will have a common platform to assess and compare different models. This will enable meaningful comparisons and help identify promising approaches for further research and development in the field.
\end{abstract}

% \keywords{benchmark, }

\maketitle

\section{Introduction}
Citation recommendation is the process of providing accurate citations for the intended cited papers. In order to accomplish this, citation recommendation systems make use of various features from the citing or cited paper. These features encompass the metadata and text content, as well as the contextual information surrounding the citation. The metadata typically includes the title, author, abstract, venue, and published date. Based on these features, citation recommendation systems can be categorized as either global or local.

Global citation recommendation systems rely on comprehensive information, including metadata and content of papers, to predict citations. For example, Bethard and Jurafsky \cite{bethard2010should} utilized only the title and abstract in their model, while ClusCite \cite{ren2014cluscite} incorporated both metadata and article content, using terms, authors, and venues to group papers based on shared interests for similar paper recommendations. Bhagavatula et al. \cite{bhagavatula2018content} and Cohan et al. \cite{cohan2020specter} leveraged the content of articles for citation recommendations.

In contrast, local citation recommendation systems focus on the contextual aspects of the citation. Some of these methods combine context with metadata \cite{gu2022local, medic2020improved, yang2018lstm, he2010context, yin2017personalized, huang2015neural}. Depending on the approach, different character limits are used, such as 100 \cite{gu2022local, medic2020improved, he2010context, ebesu2017neural, jeong2020context}, 200 \cite{medic2020improved}, 600 \cite{yang2018lstm, yin2017personalized}, or 2048 \cite{taylor2022galactica}. Additionally, the tokens considered for the context can be either characters \cite{medic2020improved, gu2022local, ebesu2017neural} or words \cite{he2010context}. The context can be bidirectional \cite{he2010context, ebesu2017neural, gu2022local, medic2020improved, huang2015neural} or unidirectional \cite{jeong2020context}.

Ongoing research in citation recommendation also emphasizes the significance of datasets utilized for training these models. The existing citation datasets are sourced from a wide range of publication fields and journals, encompassing diverse meta-information extracted from the papers. Specific datasets cater to particular domains, such as ACL-ARC \cite{bird2008acl} which focuses on data from the computational linguistics, DBLP \cite{tang2008arnetminer} which comprises datasets from the field of computer science, PubMed which contains data from the biomedical domain, and FullTextPeerRead \cite{jeong2020context} which consists of datasets from ACL and PeerRead. On the other hand, datasets like Refseer \cite{huang2014refseer}, Citeseer \cite{giles1998citeseer}, S2ORC \cite{lo2019s2orc}, Citeomatic Opencorpus \cite{bhagavatula2018content}, and CORE \cite{knoth2012core} are derived from different fields. Similarly, datasets such as unArxiv \cite{saier2020unarxive} and Bibliometric-Enhanced arXiv \cite{saier2019bibliometric} include papers extracted from Arxiv.

Different models employ diverse dataset parameters to accommodate their unique structural requirements, often necessitating the need for filtering and format changes to align with their respective models. In addition, authors frequently construct their own datasets tailored to their specific citation recommendation methods \cite{bhagavatula2018content, huang2014refseer, jeong2020context}. However, the creation of a standardized dataset presents challenges due to the variations in formats and structures that are tailored to individual models. The diversity in models, datasets, and evaluation metrics poses challenges when assessing and comparing citation recommendation approaches~\cite{ali2021overview}. To address this issue, there is a pressing need for a standardized dataset and evaluation metrics that can be used to evaluate these models consistently.

This work proposes developing a benchmark for analyzing and comparing citation recommendation models. In our experiment, our primary focus will be developing a benchmark for local citation recommendation systems. This benchmark will assess the performance of models on distinct features of the citation context and provide a comprehensive evaluation of the citation recommendation models across all these features, presenting the results in a standardized manner. By establishing a benchmark with standardized evaluation metrics, researchers and practitioners in the field of citation recommendation will have a common platform for assessing and comparing different models. This will enable efficient evaluation of citation recommendation models across various domains and contexts, eventually guiding to improved model performance and a more in-depth understanding of citation recommendation systems.

The contributions of this work are as follows:

\begin{itemize}
  \item The aim is to curate a range of diagnostic datasets, each focusing on different aspects of the citing text and metadata of the cited paper. In the experiment, the primary focus is on the development of a benchmark for local citation recommendation systems. The existing datasets S2ORC \cite{lo2019s2orc} and S2AG \cite{kinney2023semantic} are used to generate diagnostic datasets. 
  
  \item These diagnostic datasets serve as the basis for evaluating the performance of various citation recommendation models. The evaluation of these models incorporates a range of metrics including Recall and Mean Reciprocal Rank (MRR), where we use BM25 as a baseline model. 

\end{itemize}

The source code, diagnostic datasets, and benchmark model for this project are available online. The source code of this project is available at github\footnote{\url{https://github.com/puzzz21/citeBench}} and the diagnostic datasets are available at google-drive\footnote{\url{https://drive.google.com/drive/folders/1JJ-Xbg-Tnh-2qeMEOFkaJahyFID8d7Og}}. 

\section{Related work}
The field of citation recommendation has seen extensive research encompassing diverse methods and evaluation techniques. These systems are categorized based on their underlying methodologies. Ali et al. \cite{ali2021overview} classified citation recommendation approaches into Collaborative Filtering (CF), Content-Based Filtering (CB), Graph-Based (GB) filtering, Deep Learning-Based (DL) models, and Hybrid Filtering. Similarly, Ko et al. \cite{ko2022survey} distinguished these techniques into text mining, KNN, clustering, matrix factorization, and neural networks. Collaborative filtering and matrix factorization utilize user ratings and preferences to suggest new items, while content-based filtering relies on the metadata of papers such as titles, abstracts, and keywords. Text mining techniques like TF-IDF analyze semantic content for recommendation, and graph-based approaches utilize relationships among entities such as authors, papers, and venues. Deep learning methods, on the other hand, employ models trained on citation contexts or full texts to predict citations. KNN and clustering methods further refine recommendations by identifying similar papers or grouping relevant features.

Hybrid models combine these approaches to achieve improved performance. For instance, He et al. \cite{he2010context} introduced a context-aware recommendation model based on local context using a non-parametric probabilistic framework, emphasizing the effectiveness of local context (citation surroundings) over global context (title and abstract). Färber et al. \cite{farber2020citation} further categorized local citation recommendation into feature-based models, topic modeling, machine translation, and neural networks, which leverage varying features, from textual similarity to neural embedding techniques.

Citation recommendation models use different context sizes and methodologies. Jeong et al. \cite{jeong2020context} demonstrated that context size significantly impacts performance, favoring a bi-directional context of approximately 100 tokens. Similarly, Medić et al. \cite{medic2020improved} found that smaller contexts, about 200 characters, yield better results. Techniques also vary, including probabilistic models \cite{he2010context}, LSTM-based models \cite{yang2018lstm}, and Graph Convolutional Networks (GCNs) coupled with BERT embeddings \cite{jeong2020context}. Many models utilize a two-phase system: a candidate selection step, often using fast retrieval techniques such as KNN or BM25, followed by a reranker phase for finer evaluation. Notable contributions in this domain include SciBERT-based reranking models \cite{nogueira2020navigation, gu2022local} and transformer architectures such as Specter \cite{cohan2020specter} and Galactica \cite{taylor2022galactica}.

Negative sampling techniques play a crucial role in model training. For instance, Gu et al. \cite{gu2022local} retrieved candidates using a hierarchical attention encoder and randomly sampled negative papers to enhance the relevance of predictions. Similarly, Nogueira et al. \cite{nogueira2020navigation} used BM25 for candidate retrieval, marking relevant papers as positive and others as negative. The careful design of negative samples, such as the use of nearest neighbors or time-based filtering, has been shown to improve performance \cite{bhagavatula2018content, wang2020content}.

Despite progress, there is no standardized benchmark for citation recommendation models, unlike other NLP domains. Deghani et al. \cite{dehghani2021benchmark} highlighted the need for cohesive benchmarks to measure progress effectively. Benchmarks in NLP, such as GLUE \cite{wang2018glue}, SuperGLUE \cite{wang2019superglue}, and GEM \cite{gehrmann2021gem}, offer insights into how comprehensive evaluation frameworks enhance task performance. These frameworks incorporate metrics like inference latency, energy efficiency, and fairness, essential for holistic assessment. For citation systems, benchmarks must address diverse phenomena, including varying citation patterns and author biases, to capture real-world applicability.

Citation patterns, such as positioning and frequency, vary significantly across disciplines, with differences in biomedical sciences, physical sciences, and humanities \cite{boyack2018characterizing}. Zhu et al. \cite{zhu2015measuring} identified that citation influence correlates with the number of in-text mentions and title similarity, while Ritchie et al. \cite{ritchie2008comparing} emphasized the importance of using broader context windows for indexing. Furthermore, newer papers (0-3 years) and older works (8+ years) often exhibit lower perceived influence compared to those published in intermediate periods.

This broad spectrum of research underscores the complexity of citation recommendation and the critical need for standardized benchmarks to ensure consistent and meaningful advancements in the field.

\section{Diagnostic Datasets}
Diagnostic datasets play a crucial role in error analysis, performance evaluation, and model comparison in citation recommendation \cite{wang2018glue}. In this work, to generate the datasets, the S2ORC dataset \cite{lo2019s2orc} is utilized, a filtered dataset obtained from the Semantic Scholar literature corpus \cite{Ammar2018ConstructionOT}. This comprehensive dataset consists of papers from medicine, physics, biology, computer science, mathematics, chemistry, materials science, psychology, engineering, environmental science, business, geography, sociology, political science, geology, economics, history, art, and philosophy. The datasets are distributed across multiple years and journals. The dataset provide comprehensive data about research papers, including their full text, annotations, author information, referenced works, and annotated content elements. The Semantic Scholar dataset S2AG \cite{kinney2023semantic} is employed to access metadata.  Diverse model-agnostic datasets are essential to ensure the practical evaluation of various models. This work aims to generate diagnostic datasets focusing on different features of the citing text extracted from the S2ORC dataset. The process includes extracting data from each field in proportion to their representation in the S2ORC dataset and focusing on specific fields as required. The approach prioritizes sentences with mid-level length and only one citation per sentence, unless specified otherwise. This method creates a diverse and representative dataset that captures various fields, years and citation count and maintains a consistent sentence structure for effective analysis and evaluation. The final dataset replaces the citation with \textit{<REF>}.
 
Below, the precisely created diagnostic datasets are described. The following datasets were combined to generate diagnostic datasets based on fields, years, and citation counts. Consequently, there are eight diagnostic datasets categorized by fields, years, citation count, context length, citation location, context type, POS of surrounding words, and low-resource.

\subsection{Datasets}

\subsubsection{Context Length}
In order to define the classes of the context length, the length of sentences in the papers must be checked. Initially, the sentences will be split at the token level, and the number of tokens in each sentence will be counted; here, the tokens represent words. The outliers in the sentence length are analyzed where, the outliers with a high number of tokens are likely due to mathematical equations and formulas that were split at the character level because of spaces between each character. These outliers will be removed using a z-score value. Any calculated z-score value greater than the assigned z-threshold value (set to 3) will be marked as an outlier and removed accordingly. Afterward, the probability distribution function will be calculated to generate a normalized distribution. The longest sentences extend up to 70 tokens. Now, sections are defined to determine the classes for this diagnostic dataset. The mean and standard deviation will be used to extract the upper and lower bounds.

\begin{figure}
    \centering
    \includegraphics[width=0.8\columnwidth]{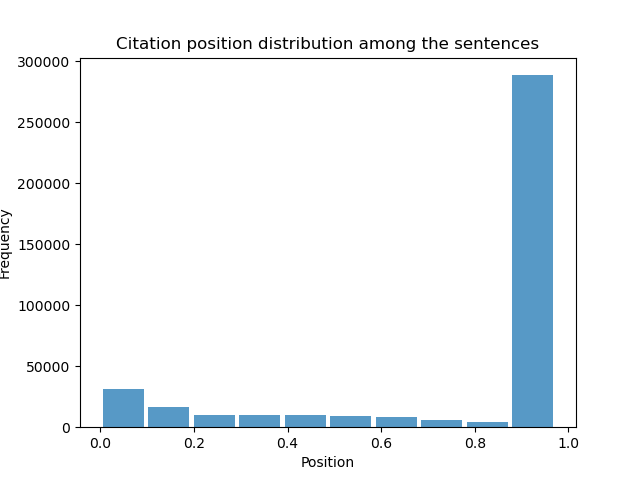}
    \caption{Citation position in the context sentences.}
    \label{fig:normalLoc}
\end{figure}

\subsubsection{Citation Location}
The location of a citation within the text can impact the performance of citation recommendation models. For example, in a study by Wright et al. \cite{wright2021citeworth}, the authors focused on the citation location at the end of a sentence to determine if it enhances clarity. First, information on the citation location of various citations in the sentence based on their token position will be collected. Initially, sentences of mid-length, as described in the previous section, will be filtered to ensure that the length distribution does not impact the results.

The position will then be normalized by dividing the position of the citation by the length of the respective sentence. Figure \ref{fig:normalLoc} shows the resulting position of the citation tokens in the sentence. The position of the citation in the sentence appears mostly at the end, with a few at the beginning.

\subsubsection{Context Type}
Citation recommendation models' performance can vary depending on the type of context. The objective is to create a diagnostic dataset encompassing diverse context types.
While different studies examine diverse types of citation contexts \cite{angrosh2010context, tacskin2018content, te2022citation, kunnath2020overview, wang2020important, zhu2015measuring}, this paper uses the citation intent of the papers defined in S2AG dataset. The focus will be on mid-length sentences with the citation location at the end.

\subsubsection{Low Resource Papers}
Many fields do not account for a sufficient proportion of the entire S2ORC dataset. Several models, such as those in \cite{bhagavatula2018content, nogueira2020navigation}, are trained on PubMed, which contains data from the medical domain, and DBLP \cite{bhagavatula2018content, yang2018lstm, cai2018three, nogueira2020navigation, thierry2023rar, ali2022spr}, which contains data from the computer science domain. Similarly, various models are trained on ACL papers which includes NLP and computational linguistic \cite{medic2020improved, gu2022local, jeong2020context, yang2018lstm, wang2020content, ali2022spr, lu2023research}. 

Citation recommendation models should also be able to provide accurate citations for less popular fields. Therefore, the aim of this diagnostic dataset is to assess how well the models perform in low-resource fields. Fields with less than 3\% of the total papers, are defined as low-resource datasets. These fields includes: psychology, environmental science, engineering, business, geography, political science, sociology, geology, history, art and philosophy.

\subsubsection{POS of Surrounding Words} In the citing text, the surrounding words of the citations can belong to different parts-of-speech (POS). For instance, in the sentence ".. is stated by Abc et al. (2022)," the citation is followed by a preposition (IN). Similarly, citations can be followed by verbs, such as ".. argues." Additionally, citations may follow proper nouns, as in "... S2ORC (Lo et al., 2019)." These surrounding words play a crucial role in determining whether the preceding or following token is the actual citation. To assess the performance of citation recommendation models in handling this aspect, a dataset will be created that includes various POS words before and after the citation.

By incorporating POS words that are less or more frequent in association with citations, the effectiveness of models in utilizing POS information for accurate citation recommendation can be evaluated. This dataset will provide insights into the model's ability to consider the context and surrounding linguistic cues when making citation predictions.

\begin{figure}
\centering
\includegraphics[width=0.8\columnwidth]{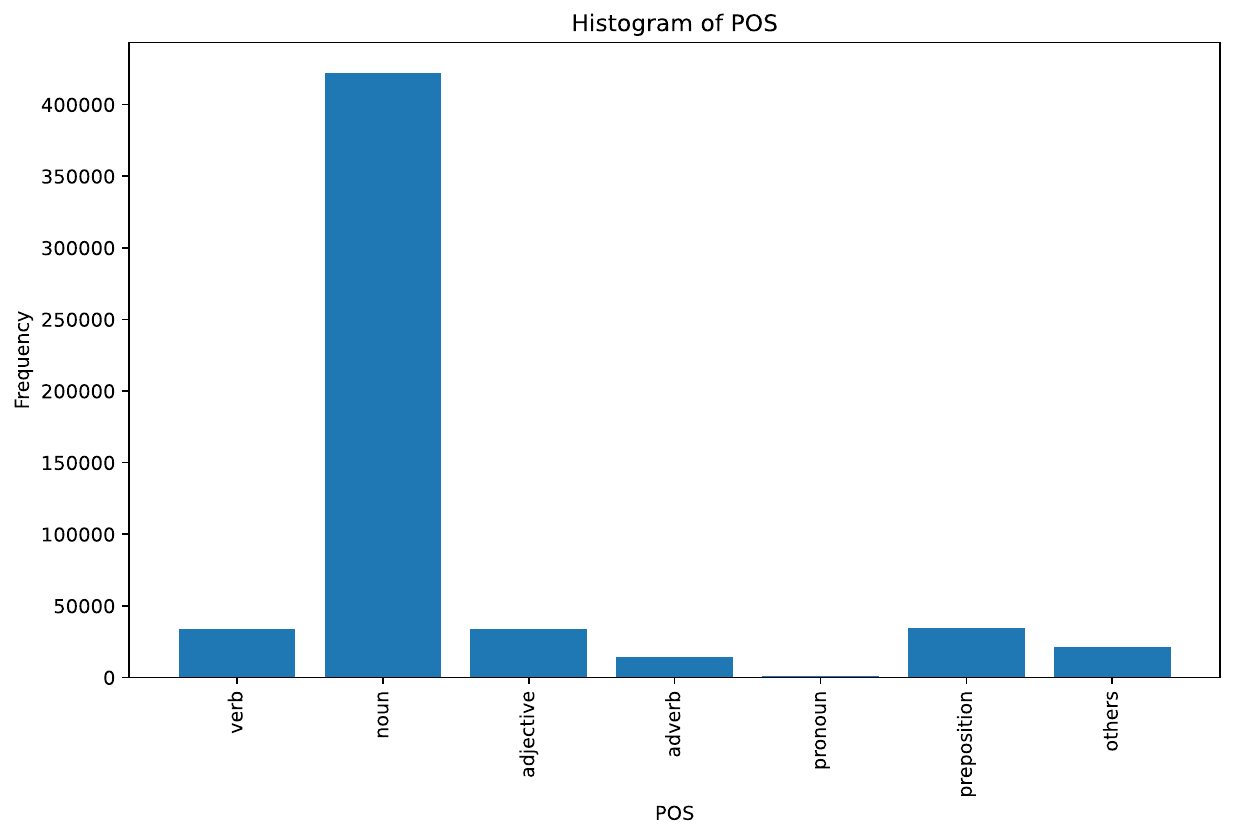}
\caption{Combined preceding POS of the citation.}
\label{pics:posnormalize}
\end{figure}

% \begin{figure}
%     \centering
%     \includegraphics[width=0.8\columnwidth]{2019_master_thesis_example/fig/visualization/POSCount2.png}
%     \caption{Combined preceding POS of the citation.}
%     \label{fig:posnormalize}
% \end{figure}

The POS of the preceding and following words in citations is analyzed to obtain the results. Figure \ref{pics:posnormalize} shows the distribution of the combined POS words by combining the POS tags as follows:

\begin{itemize}
\item verb : VB, VBD, VBG, VBP, VBZ, VBN
\item noun : NN, NNS, NNP, NNPS
\item adjective : JJ, JJR, JJS
\item adverb : RB, RBR, RBS
\item pronoun : PRP, PRP\$, WP, EX
\item preposition : IN, TO
\item others : PDT, DT, CC, MD, WDT, POS, RP, FW, CD, WRB

\end{itemize}

The combined POS tags in the dataset are utilized. The exact process is repeated for the following POS as well. The distribution is examined to identify common POS patterns. Subsequently, these POS categories are utilized to differentiate the POS tags of the preceding and following words. Classes are created based on the preceding and following POS tags, with sub-classes for the specific POS tags within each class. A subsequent manual check ensures the correctness of the dataset.

\subsection{Data Extraction}

\begin{figure}
    \centering
    \includegraphics[width=1.1\columnwidth]{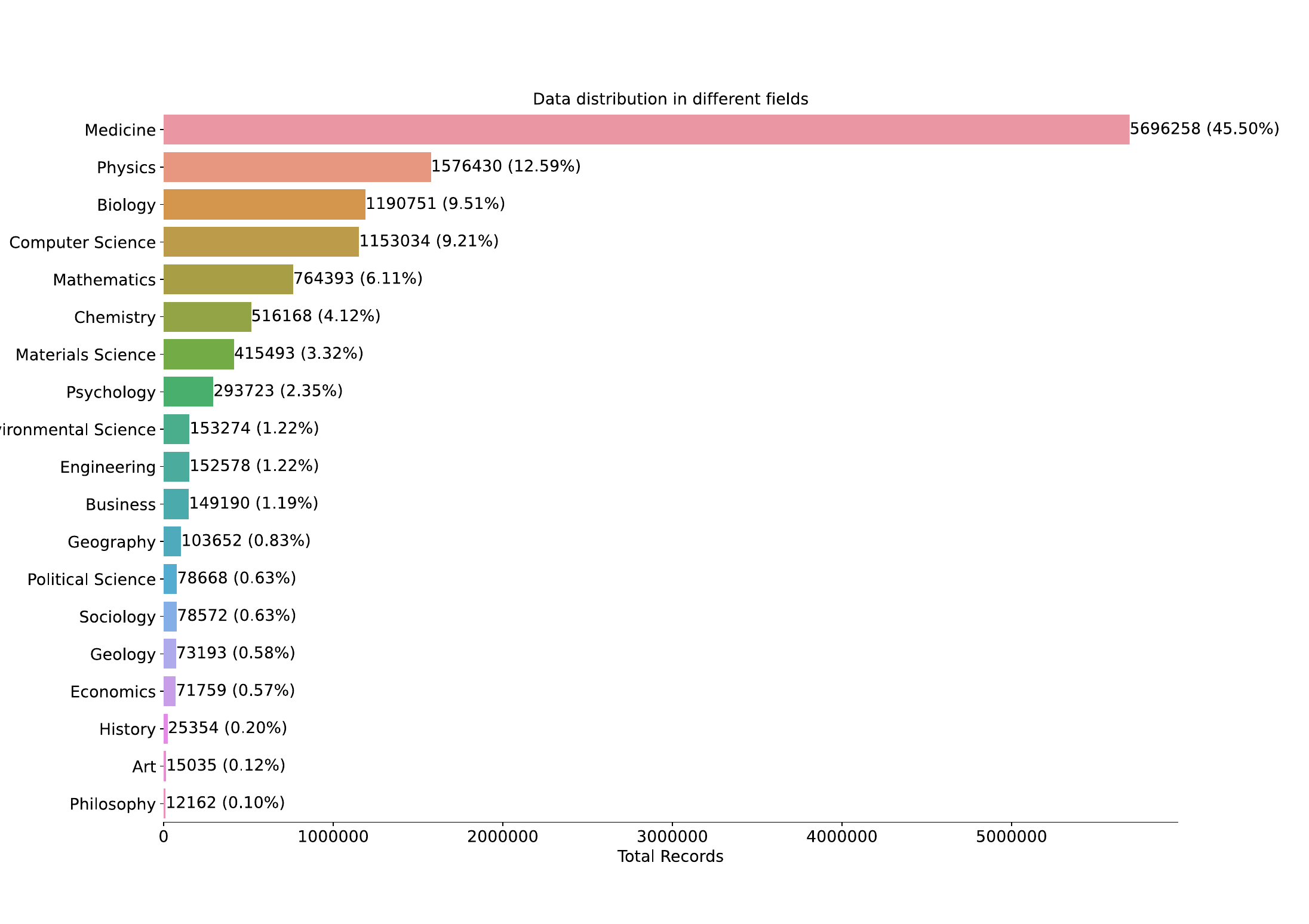}
    \caption{Data distribution of papers according to various fields}
    \label{fig:DSDist}
\end{figure}

\begin{figure}
    \centering
    \includegraphics[width=1.1\columnwidth]{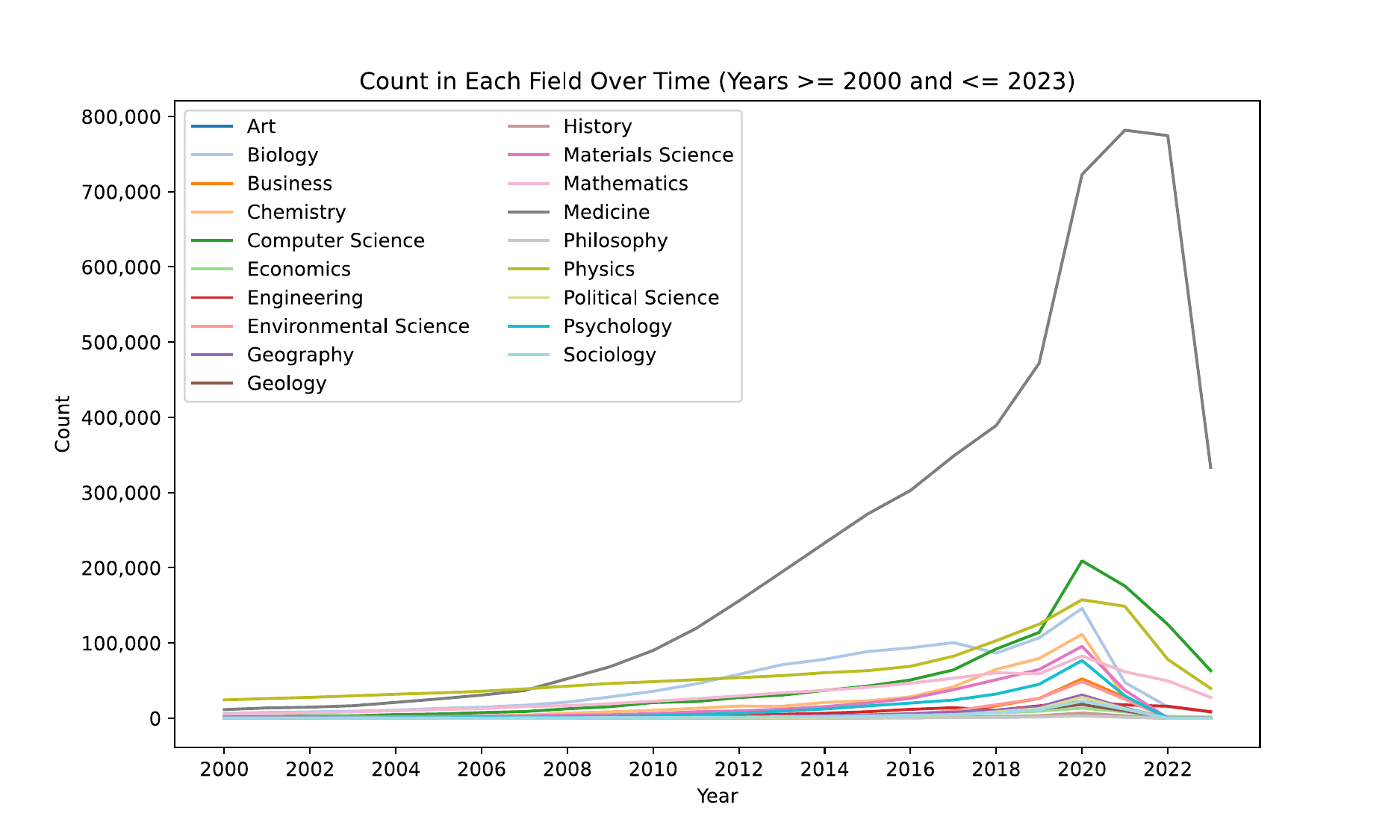}
    \caption{Data distribution of papers, from years 2000 to 2023 with respect to fields}
    \label{fig:DSYrDist}
\end{figure}

\begin{figure}
    \centering
    \includegraphics[width=1.08\columnwidth]{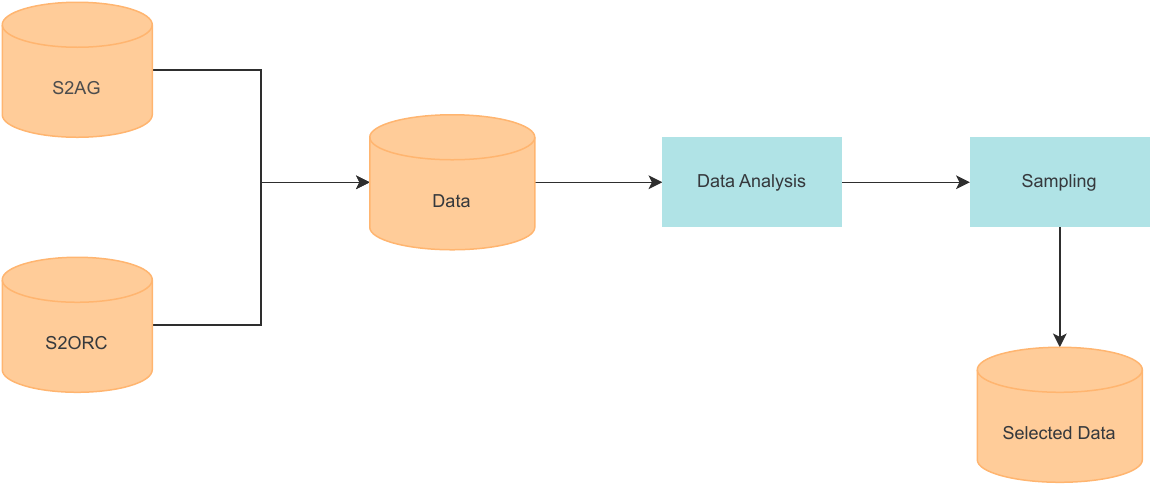}
    \caption{Process of data selection}
    \label{fig:dataStorage}
\end{figure}

 To collect the samples, the dataset is analyzed first, and samples are collected accordingly, as depicted in Figure \ref{fig:dataStorage}. Only in the first stage are the papers sampled, and in the second stage, the citing sentences are extracted according to the requirements of the diagnostic datasets.
Upon obtaining the S2ORC data alongside its metadata from S2AG, the next step involves extracting fields, years, and citation count information from these datasets. The data is then analyzed based on these parameters. As depicted in Figure \ref{fig:DSDist}, the initial analysis focuses on the total records within respective fields.
The most prominent publications within our dataset fall within the field of medicine, constituting 45.50\% of the entire dataset. In contrast, history, art, and philosophy together account for less than 0.50\% of the total dataset. Figure \ref{fig:DSYrDist} illustrates the total papers in S2ORC from the years 2000 to 2023, categorized by each field. Notably, the dataset demonstrates an increase in papers until 2020, followed by a decline. Medicine contains a substantial number of papers, followed by computer science, physics, and biology.

Similarly, the distribution of citation counts across all fields will be examined. Given the varying record counts in different fields, the average citation count in each field will be calculated by normalizing the sum of citation counts in each field based on the total number of records within that field. The $m$ fields can be labeled as $f_1, f_2, \ldots, f_m$. The average citation count in the $i$-th field for a total of $n$ papers is depicted in the mathematical expression below, where $C_{avg}(f_i)$ is the average citation count in field $i$.

\begin{equation}
{C_{avg}}({f_i}) = \frac{\sum_{j=1}^{n_i} C_{ij}}{n_i}
\end{equation}

Here, $C_{ij}$ is the citation count of the $j$-th paper in the $i$-th field, and $n_i$ is the total number of papers in the $i$-th field.

\begin{figure}
    \centering
    \includegraphics[width=1\columnwidth]{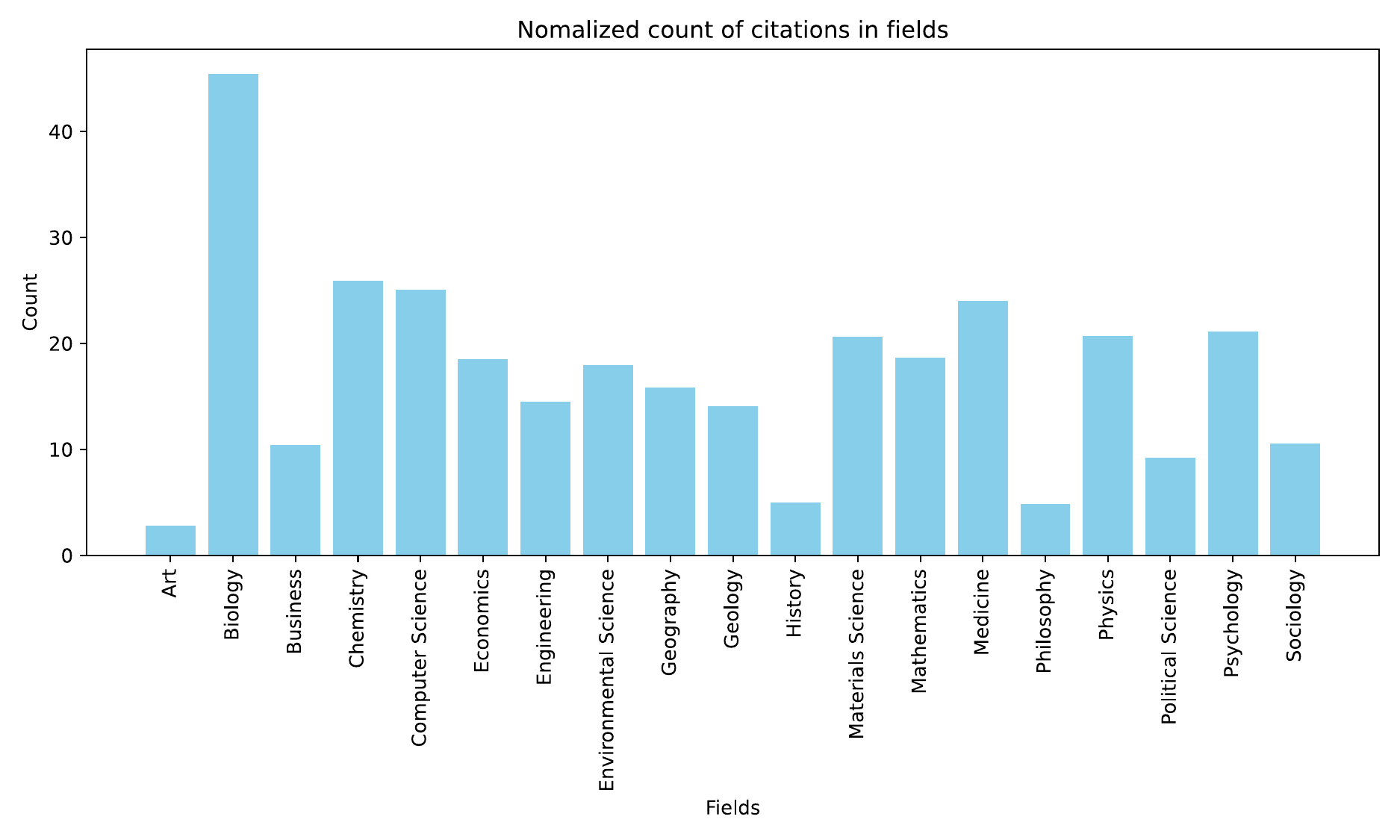}
    \caption{Citation count distribution based on fields, where the citation count value is normalized based on total papers in each fields.}
    \label{fig:DSCCDist}
\end{figure}

The resulting value $C_{avg}({f_i})$ is presented in Figure \ref{fig:DSCCDist} for all the fields $m$, with biology exhibiting the highest average citation count. Having conducted this data analysis, it becomes evident that the distribution of papers significantly varies across fields and their corresponding citation counts.

\subsubsection{Paper Selection}
\begin{figure*}
    \centering
    \includegraphics[width=2\columnwidth]{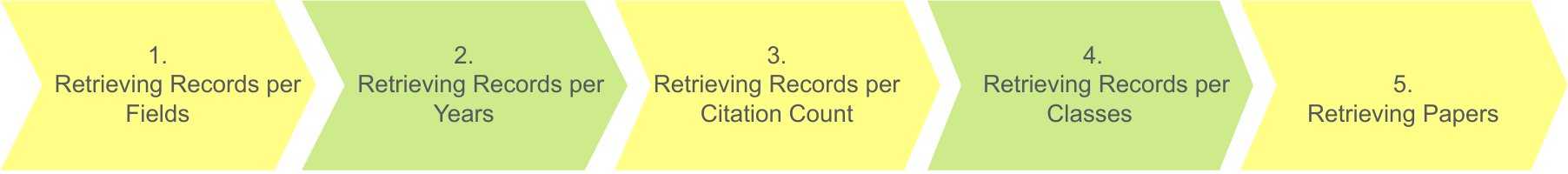}
    \caption{Process of paper selection}
    \label{fig:flow1}
\end{figure*}

In this section, the approach for paper selection is described. The processes followed are illustrated in Figure \ref{fig:flow1}.

For the sampling process, fields, years, and citation counts are used as parameters. This decision is based on the premise that citation recommendation models should be capable of providing accurate citations regardless of the paper's field, publication year, and the popularity of the cited paper. By sampling data from each of these strata, the goal is to create the diagnostic datasets that is evenly distributed.

Our sampling strategy encompasses a sequence of steps and implements a hierarchical approach of stratified sampling similar to the one described in the work by Trost (1986) \cite{trost1986statistically}. This hierarchical strategy comprises multiple tiers: initially, papers are categorized according to their respective fields; subsequently, within each field, papers are segregated into specific year groups; ultimately, within each year group, further divisions are made based on the citation count of the papers. Moreover, these citation count groups are subsequently partitioned in accordance with the classes defined within the diagnostic datasets. For instance, within the context of the citation length dataset, the citation count groups are divided into three categories: long, medium, and short context lengths.

The sampling processes for all the diagnostic datasets follow the processes described below. The steps for retrieving the data from cited papers are defined in the following sections.

\begin{enumerate}

\item{Retrieving Records per Fields}

To initiate the process, the data from each field within the dataset is extracted. Here, the data is not uniformly distributed across different strata. To achieve an equal distribution of records within these strata, it becomes necessary to exclude fields whose data accounts for less than 3\% of the total dataset. Consequently, the following fields—chemistry, mathematics, computer science, material science, biology, physics, and medicine—are retained due to their substantial record counts. These fields are used for all the diagnostic datasets, with the exception of the low-resource diagnostic dataset, for which papers are retrieved from fields accounting for less than 3\% of the total data in order to assess how the model performs. For low-resource datasets, records are retrieved from fields such as psychology, environmental science, engineering, business, geography, political science, sociology, geology, history, art, and philosophy.
% These datasets will serve the purpose of evaluating the performance of citation recommendation models when dealing with datasets that are less prevalent in real-world scenarios.

\item{Retrieving Records per Years}

The records from the specified fields are filtered based on publication years. Within each field, records are selected based on their respective publication years, categorizing these years into five distinct groups: 1996-2000, 2001-2005, 2006-2010, 2011-2015, and 2016-2020. This division into five-year increments is driven by the availability of records suitable for our sampling approach.

\item{Retrieving Records per Citation Count}

For each designated group of sampled years, the citation counts for the papers within that specific group are collected. Subsequently, these citation counts are organized in ascending order. Following this, the citation counts are divided by 10, ensuring that the resulting ranges of citation counts encompass an equal number of records within each group. Unique records are then sampled across all diagnostic datasets.

In this context, the selection of citation counts begins from 5. This approach is rooted in the fact that cited papers are sampled with the objective of extracting pertinent context from the relevant citing papers. It's important to acknowledge that these contexts might occasionally not align precisely with the defined criteria of the diagnostic dataset classes. To account for this variability and ensure the availability of multiple options that meet context requirements, cited papers with more than 4 citations are sampled. This methodology increases the chances of acquiring suitable contexts that align with the desired criteria.

\item{Retrieving Records per Classes}

As the papers are sampled based on the requirements of each diagnostic dataset, the number of samples needed is defined first. Not all sampled papers will make it into the diagnostic dataset, as there are certain conditions that need to be fulfilled. Therefore, an upper limit value for sampling the papers is established. For example, for the diagnostic dataset regarding "length," 10 papers are sampled for each group. Although the diagnostic dataset for length requires only 3 records, the upper limit defined is 10.

\item{Retrieving Papers}

Based on the above-retrieved data, the metadata of these papers is stored in S2AG within a collection in MongoDB. The models are designed to predict these papers given the citing context. To retrieve the citing context, papers from the corpus that cite the collected cited papers are obtained. All citing papers of the collected cited papers are retrieved, and their metadata is saved from S2AG in the collection. These papers should be open source and include all the required metadata, such as author, title, journal, venue, and publication date. Subsequently, the respective text of the citing papers is retrieved from the S2ORC dataset. The data is then saved in different collections in the database.

\end{enumerate}

\subsubsection{Sentence Extraction}

\begin{figure*}
    \centering
    \includegraphics[width=2\columnwidth]{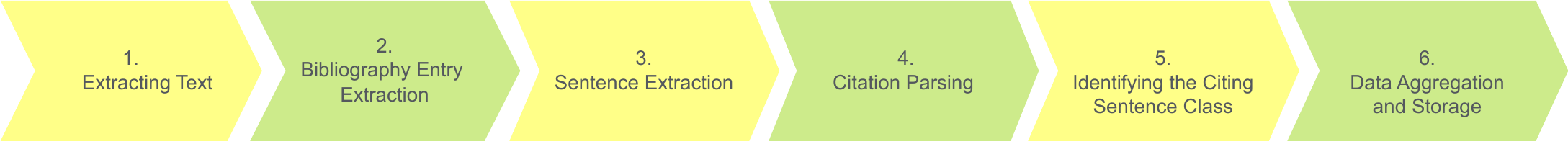}
    \caption{Process of sentence extraction}
    \label{fig:flow2}
\end{figure*}

We will follow the steps defined below for the extraction of the citing sentences, as shown in Figure \ref{fig:flow2}.

\begin{enumerate}

\item{Extracting Text}

First, the citing and cited papers needs to be collected. There are three collections for each dataset, consisting of the cited and citing paper's metadata and the citing paper's text. The citing papers for each cited paper are retrieved from the MongoDB collection for the specified dataset. Then, the text content of the citing paper is extracted. The text content of the papers is retrieved by parsing each annotated paragraph defined in the S2ORC dataset.

\item{Bibliography Entry Extraction}

In addition to analyzing sentences, a comprehensive analysis of references within each sentence will be conducted. The focus is on retrieving relevant reference information and citation text. The S2ORC dataset provides annotations for cited bibliographies and their respective citations in the text. To accurately parse references, specific attributes within the sentence will be examined, including bibliography references, text citations, cited paper authors, cited paper titles, and cited paper publication venues, to ensure their presence in S2ORC.

As mentioned earlier, the S2ORC dataset contains bibliography entries along with bibliography references and links them through a unique identifier. The bibliography references will be extracted using the corpus ID of the cited paper. Once these references are obtained, the unique identifier will be used to identify the bibliography entry in the text content, such as [12] or (abc et al., 2012), etc. This value will be used in the further process of citing sentences for the respective cited paper.

\item{Sentence Extraction}

Next, individual sentences will be extracted from the retrieved paragraphs using the SciSpacy\footnote{\url{https://allenai.github.io/scispacy}} library and the SciBERT\footnote{\url{https://huggingface.co/allenai/scibert_scivocab_uncased}} model. To ensure sentence integrity, each sentence will be verified to conclude with appropriate punctuation (period, question mark, or exclamation mark), trimmed as necessary, and a space will be added at the end. Additionally, sentences containing footer notes with superscripts (e.g., "7 this is...") will be identified and excluded by checking the first character of the sentence. Sentences that contain only the citation entry of the respective paper retrieved from the above process will be returned. Sentences with only one citation entry will be retrieved since the diagnostic datasets consist of sentences containing only one citation. The sentences will be filtered accordingly.

\item{Citation Parsing}

Instances in the S2ORC dataset reveal citations that do not adhere to standard citation rules. For example, citations such as "(Lee et al., 1997a, Lee et al., 1997b)" are incorrectly transformed into "(Lee et al., 1997a(Lee et al., 1997b)". To tackle this issue, regular expressions are utilized to identify the standard citation structures of various citation styles. A citation parser is developed to determine if a sentence contains correct citations following various citation styles and to retrieve the citation text. The citation parser for the initial cited sentence selection checks for one correct citation in the sentence. Subsequently, the citation text within the sentence is replaced with the placeholder "<REF>" and the replaced citation is added to the metadata for future reference.

\item{Identifying the Citing Sentence Class}

Each diagnostic dataset represents various features of the citing sentences. These parameters are defined by different classes of these features. The classes and features of the citing sentences for the diagnostic datasets are defined below:

\begin{enumerate}
\item Fields: Biology, Chemistry, Computer Science, Medicine,
Materials Science, Physics, Mathematics
\item Years: 1996-2000, 2001-2005, 2006-2010, 2011-2015, 2016-2020
\item Citation Count Groups: 0, 1, 2, 3, 4, 5, 6, 7, 8, 9
  \item Context Length: Short, Medium, Long
  \item Context Location: First, Middle, Last
  \item Context Type: Background, Result, Method
  \item POS of Surrounding Words:
    \begin{enumerate}
        \item Preceding POS: verb, noun, adjective, adverb, pronoun, preposition, and others
        \item Following POS: verb, noun, adjective, adverb, pronoun, preposition, and others
    \end{enumerate}
  \item Low Resource: Psychology, Environmental Science,
  Engineering, Business, Geography, Political Science,
  Sociology, Geology, History, Art, Philosophy
\end{enumerate}

Once the classes are retrieved, the sentence is filtered based on the mentioned class. If it complies, additional conditions are applied accordingly. These conditions ensure that other features, such as length or position, do not affect the evaluation result. The additional conditions applied to each dataset are listed below:

\begin{enumerate}
  \item Context Length: Reference at the "end" of the sentence
  \item Context Location: The length of the citing sentence belongs to the class "medium"
  \item Context Type: The length of the citing sentence belongs to the class "medium" and the reference is at the "end" of the sentence
  \item POS of Surrounding Words: The length of the citing sentence belongs to the class "medium" and the reference is at the "middle" of the sentence
  \item Low Resource: The length of the citing sentence belongs to the class "medium" and the reference is at the "end" of the sentence
\end{enumerate}

\item{Data Aggregation and Storage}

The cited sentences that pass the mentioned conditions in the previous steps are stored in MongoDB. The stored information includes the field, year, citation count group, cited paper corpus ID, citing paper corpus ID, citing sentence where the reference is replaced by "<REF>", the citation entry, and the calculated class according to each diagnostic dataset. The paper's metadata comprises fields of study, venues, open-access status, publication year, citation count, influential count, journal, publication date, external IDs, reference count, and publication date. Similarly, the author's metadata includes the author's name, aliases, affiliations, homepage, paper count, citation count, and h-index, which can be retrieved using the script with the paper's corpus ID if necessary.

\end{enumerate}

\begin{table*}[htbp]
    \centering
    \resizebox{\textwidth}{!}{%
    \begin{tabular}{lccccccccccccl}
        \toprule
        \textbf{Dataset} & \textbf{Classes} & \multicolumn{2}{c}{\textbf{NCN}} & \multicolumn{2}{c}{\textbf{LCR}} & \multicolumn{2}{c}{\textbf{Galactica 125m}} & \multicolumn{2}{c}{\textbf{Galactica 1.3b}} & \multicolumn{2}{c}{\textbf{Galactica 6.7b}} & \multicolumn{2}{c}{\textbf{BM25}} \\
        \cmidrule(lr){3-4}\cmidrule(lr){5-6}\cmidrule(lr){7-8}\cmidrule(lr){9-10}\cmidrule(lr){11-12}\cmidrule(lr){13-14}
        &  & R@10 & M@10 & R@10 & M@10 & R@10 & M@10 & R@10 & M@10 & R@10 & M@10 & R@10 & M@10 \\
        \midrule
        % Sample row formatting (add data rows here)

\multirow{7}{*}{Fields}  & Biology & 0.0126 & 0.0040 & 0.0013 & 0.0006 & 0.0 & 0.0 & 0.0226 & 0.0105 & 0.0327 & 0.0136 & \textbf{0.195} & \textbf{0.1072} \\
 & Chemistry & 0.0115 & 0.0035 &  0.0 & 0.0 & 0.0 & 0.0 & 0.0033 & 0.0008 & 0.0115 & 0.0031 & \textbf{0.2158} & \textbf{0.1183} \\
 & Computer Science & 0.0221 & 0.0061 & 0.0063 & 0.0 & 0.0063 & 0.0015 & 0.0505 & 0.0236 & 0.0410 & 0.0209 & \textbf{0.194} & \textbf{0.0949} \\
  & Medicine & 0.0081 & 0.0019 & 0.0049 & 0.0 & 0.0016 & 0.0016 & 0.0309 & 0.0148 & 0.0276 & 0.0097 & \textbf{0.187} & \textbf{0.0963} \\
 & Materials Science & 0.0109 & 0.0026 & 0.0109 & 0.0 & 0.0 & 0.0 & 0.0 & 0.0 & 0.0088 & 0.0015 & \textbf{0.1969} & \textbf{0.0986} \\
 & Physics & 0.0222 & 0.0077 & 0.0154 & 0.0 & 0.0 & 0.0 & 0.0171 & 0.0075 & 0.0188 & 0.0081 & \textbf{0.1094} & \textbf{0.0488} \\
 & Mathematics & 0.0299 & 0.0103 & 0.0456 & 0.0001 & 0.0 & 0.0 & 0.0173 & 0.0074 & 0.0220 & 0.0087 & \textbf{0.1714} & \textbf{0.082} 

 \\\midrule

 \multirow{5}{*}{Years}  & 1996-2000 & 0.0211 & 0.0050 & 0.0179 & 0.0006 & 0.0 & 0.0 & 0.0042 & 0.0016 & 0.0105 & 0.0042 & \textbf{0.1568} & \textbf{0.0757}  \\
 & 2001-2005 & 0.0202 & 0.0061 & 0.0192 & 0.0 & 0.0 & 0.0 & 0.0125 & 0.0058 & 0.0202 & 0.0078 & \textbf{0.19} & \textbf{0.0942} \\
 & 2006-2010 & 0.0219 & 0.0075 & 0.0079 & 0.0 & 0.0009 & 0.0002 & 0.0219 & 0.0119 & 0.0219 & 0.0100 & \textbf{0.2112} & \textbf{0.1115}  \\
  & 2011-2015 & 0.0121 & 0.0044 & 0.0095 & 0.0 & 0.0017 & 0.0010 & 0.0242 & 0.0109 & 0.0260 & 0.0089 & \textbf{0.2016} & \textbf{0.1015}  \\
 & 2016-2020 & 0.0143 & 0.0045 & 0.0009 & 0.0 & 0.0018 & 0.0004 & 0.0250 & 0.0097 & 0.0330 & 0.0128 & \textbf{0.195} & \textbf{0.1071}

 \\\midrule

 \multirow{10}{*}{Citation Count}  & 0 & 0.0079 & 0.0066 & 0.0421 & 0.0221 & 0.0026 & 0.0007 & 0.0053 & 0.0035 & 0.0026 & 0.0013 & \textbf{0.3816} & \textbf{0.1941}  \\
 & 1 & 0.0102 & 0.0031 & 0.0305 & 0.0215 & 0.0 & 0.0 & 0.0041 & 0.0007 & 0.0102 & 0.0019 & \textbf{0.2724} & \textbf{0.1384}  \\
 & 2 & 0.0130 & 0.0037 & 0.0148 & 0.0092 & 0.0 & 0.0 & 0.0037 & 0.0011 & 0.0185 & 0.0064 & \textbf{0.2074} & \textbf{0.1169}  \\
  & 3 & 0.0111 & 0.0037 & 0.0111 & 0.0043 & 0.0 & 0.0 & 0.0111 & 0.0040 & 0.0166 & 0.0031 & \textbf{0.1937} & \textbf{0.1058}  \\
   & 4 & 0.0121 & 0.0033 & 0.0069 & 0.0017 & 0.0 & 0.0 & 0.0121 & 0.0062 & 0.0155 & 0.0054 & \textbf{0.1672} & \textbf{0.0875}  \\
    & 5 & 0.0203 & 0.0056 & 0.0034 & 0.0006 & 0.0 & 0.0 & 0.0135 & 0.0042 & 0.0203 & 0.0076 & \textbf{0.1872} & \textbf{0.1005}  \\
     & 6 & 0.0138 & 0.0030 & 0.0086 & 0.0027 & 0.0 & 0.0 & 0.0241 & 0.0078 & 0.0241 & 0.0069 & \textbf{0.1621} & \textbf{0.0813}  \\
      & 7 & 0.0207 & 0.0050 & 0.0 & 0.0 & 0.0 & 0.0 & 0.0207 & 0.0110 & 0.0155 & 0.0062 & \textbf{0.1448} & \textbf{0.0743}  \\
       & 8 & 0.0294 & 0.0110 & 0.0017 & 0.0003 & 0.0 & 0.0 & 0.0242 & 0.0131 & 0.0363 & 00170 & \textbf{0.1537} & \textbf{0.0657}  \\
        & 9 & 0.0348 & 0.0100 & 0.0018 & 0.0006 & 0.0073 & 0.0031 & 0.0568 & 0.0286 & 0.0604 & 0.0299 & \textbf{0.1264} & \textbf{0.0621}

 \\\midrule

\multirow{3}{*}{Length}  & Short & 0.0186 & 0.0041 & \textbf{0.2538} & 0.0042 & 0.0 & 0.0 & 0.0433 & 0.0204 & 0.0528 & 0.0219 & 0.1486 & \textbf{0.0764}  \\
 & Medium & 0.0095 & 0.0041 &  \textbf{0.2511} & 0.0099 & 0.0031 & 0.0008 & 0.0285 & 0.0171 & 0.0411 & 0.0166 & 0.1709 & \textbf{0.0996}  \\
 & Long & 0.0127 & 0.0020 &  \textbf{0.2518} & 0.0127 & 0.0031 & 0.0008 & 0.0411 & 0.0188 & 0.0443 & 0.0187 & 0.1677 & \textbf{0.0952}
 
 \\\midrule

\multirow{3}{*}{Location}  & Start & 0.0154 & 0.0070 & 0.2518 & 0.0151 &  0.0 & 0.0 &  0.0 & 0.0 & 0.0 & 0.0 & \textbf{0.2646} & \textbf{0.1594} \\
 & Middle & 0.0304 & 0.0101 & \textbf{0.2507} & 0.0011 & 0.0 & 0.0 &  0.0 & 0.0 & 0.0 & 0.0 & 0.2219 & \textbf{0.1042} \\
 & End & 0.0121 & 0.0019 & \textbf{0.2537} & 0.0085 & 0.0060 & 0.0037 &  0.0332 & 0.0147 & 0.0333 & 0.0108 & 0.1873 & \textbf{0.1032}
 \\\midrule

\multirow{3}{*}{Intent}  & Background & 0.0093 & 0.0026 & \textbf{0.284} & 0.0052 & 0.0 & 0.0 & 0.0469 & 0.0214 & 0.0467 & 0.0225 & 0.1838 & \textbf{0.0996}  \\
 & Methodology & 0.0129 & 0.0047 & \textbf{0.2762} & 0.0 & 0.0032 & 0.0006 & 0.0675 & 0.0293 & 0.0804 & 0.0340 & 0.1833 & \textbf{0.0935}  \\
 & Result & 0.0164 & 0.0077 & \textbf{0.1834} & 0.0 & 0.0 & 0.0 & 0.0273 & 0.0099 & 0.0437 & 0.0151 & 0.1421 & \textbf{0.0714}
 \\\midrule

\multirow{11}{*}{Low Resource}  & Psychology & 0.0 & 0.0 & 0.1166 & 0.0 & 0.0 & 0.0 & 0.0070 & 0.0008 & 0.0280 & 0.0079 & \textbf{0.2587} & \textbf{0.1277}  \\
 & Environmental Science & 0.0082 & 0.0012 & 0.1012 & 0.0 & 0.0 & 0.0 & 0.0082 & 0.0027 & 0.0246 & 0.0053 & \textbf{0.1967} & \textbf{0.122}  \\
 & Engineering & 0.0273 & 0.0058 & 0.0933 & 0.0121 & 0.0 & 0.0 & 0.0182 & 0.0109 & 0.0273 & 0.0080 & \textbf{0.2273} & \textbf{0.1068}  \\
  & Business & 0.0317 & 0.0145 & 0.1044 & 0.0013 & 0.0 & 0.0 & 0.0079 & 0.0040 & 0.0079 & 0.0040 & \textbf{0.1746} & \textbf{0.0891}  \\
 & Geography & 0.0144 & 0.0082 & 0.1137 & 0.0 & 0.0 & 0.0 & 0.0072 & 0.0036 & 0.0216 & 0.0022 & \textbf{0.3094} & \textbf{0.1728}  \\
 & Political Science & 0.0 & 0.0 & 0.0783 & 0.0 & 0.0 & 0.0 & 0.0 & 0.0 & 0.0 & 0.0 & \textbf{0.1848} & \textbf{0.123}  \\
 & Sociology & 0.0364 & 0.0161 & 0.0922 & 0.0 & 0.0 & 0.0 & 0.0 & 0.0 & 0.0091 & 0.0045 & \textbf{0.1909} & \textbf{0.101}  \\
 & Geology & 0.0215 & 0.0027 & 0.0796 & 0.0066 & 0.0 & 0.0 & 0.0 & 0.0 & 0.0108 & 0.0013 & \textbf{0.2473} & \textbf{0.1045}  \\
 & History & 0.0405 & 0.0098 & 0.064 & 0.0 & 0.0 & 0.0 & 0.0 & 0.0 & 0.0135 & 0.0135 & \textbf{0.2838} & \textbf{0.1459}  \\
 & Art  & 0.0488 & 0.0116 & 0.0365 & 0.0 & 0.0 & 0.0 & 0.0 & 0.0 & 0.0238 & 0.0048 & \textbf{0.3171} & \textbf{0.1583}  \\
 & Philosophy & 0.0606 & 0.0074 & 0.0297 & 0.0061 & 0.0 & 0.0 & 0.0 & 0.0 & 0.0 & 0.0 & \textbf{0.2121} & \textbf{0.0965}

 \\\midrule

\multirow{14}{*}{POS}  & Before Verb & 0.0515 & 0.0126 & 0.058 & 0.0 & 0.0 & 0.0 & 0.0 & 0.0 & 0.0 & 0.0 & \textbf{0.1649} & \textbf{0.0726}  \\
 & Before Noun & 0.0198 & 0.0068 & \textbf{0.3925} & 0.0029 & 0.0 & 0.0 & 0.0001 & 0.0005 & 0.0 & 0.0 & 0.1779 & \textbf{0.0784}  \\
 & Before Adjective & 0.0192 & 0.0036 & 0.0914 & 0.0141 & 0.0 & 0.0 & 0.0 & 0.0 & 0.0 & 0.0 & \textbf{0.1923} & \textbf{0.1004}  \\
  & Before Adverb & 0.0303 & 0.0043 & 0.0205 & 0.0 & 0.0 & 0.0 & 0.0 & 0.0 & 0.0 & 0.0 & \textbf{0.0606} & \textbf{0.0253}  \\
 & Before Pronoun & 0.0 & 0.0 & 0.0019 & 0.0 & 0.0 & 0.0 & 0.0 & 0.0 & 0.0 & 0.0 & \textbf{0.3333} & \textbf{0.0833}  \\
 & Before Preposition & 0.0049 & 0.0010 & 0.1169 & 0.0148 & 0.0 & 0.0 & 0.0146 & 0.0044 & 0.0098 & 0.0020 & \textbf{0.1512} & \textbf{0.0762}  \\
 & Before Others & 0.0147 & 0.0049 & 0.042 & 0.0147 & 0.0 & 0.0 & 0.0 & 0.0 & 0.0 & 0.0 & \textbf{0.1324} & \textbf{0.0419}  \\
 & After Verb & 0.0259 & 0.0064 & 0.1818 & 0.004 & 0.0 & 0.0 & 0.0029 & 0.0014 & 0.0 & 0.0 & \textbf{0.227} & \textbf{0.1238}  \\
 & After Noun & 0.0206 & 0.0085 & 0.11 & 0.0023 & 0.0 & 0.0 & 0.0 & 0.0 & 0.0 & 0.0 & \textbf{0.1649} & \textbf{0.0703}  \\
 & After Adjective & 0.0270 & 0.0090 & 0.023 & 0.0 & 0.0 & 0.0 & 0.0 & 0.0 & 0.0 & 0.0 & \textbf{0.0811} & \textbf{0.0444}  \\
 & After Pronoun & 0.0098 & 0.0016 & 0.049 & 0.0125 & 0.0 & 0.0 & 0.0 & 0.0 & 0.0 & 0.0 & \textbf{0.1373} & \textbf{0.0545}  \\
 & After Adverb & 0.0 & 0.0 & 0.0616 & 0.0123 & 0.0 & 0.0 & 0.0 & 0.0 & 0.0 & 0.0 & \textbf{0.125} & \textbf{0.0542}  \\
 & After Preposition & 0.0258 & 0.0058 & \textbf{0.1832} & 0.0093 & 0.0 & 0.0 & 0.0057 & 0.0021 & 0.0029 & 0.0007 & 0.1777 & \textbf{0.0744}  \\
 & After Others & 0.0151 & 0.0064 & \textbf{0.2286} & 0.004 & 0.0 & 0.0 & 0.0022 & 0.0003 & 0.0022 & 0.0004 & 0.1487 & \textbf{0.0591}
 \\\bottomrule

        % Add remaining rows here
    \end{tabular}%
    }
    \caption{Performance Comparison Across Different Criteria}
    \label{tab:comparison}

\end{table*}

\section{Result and discussion}
In this section, the results from different models on the diagnostic datasets are compared. For the evaluation, we chose Neural Citation Recommendation (NCN) \cite{ebesu2017neural}, Local Citation Recommendation with Hierarchical-Attention Text Encoder and SciBERT-Based Reranking (LCR) \cite{gu2022local}, 
Galactica \cite{taylor2022galactica} and BM25 ranking algorithm which is our baseline. First, we clarify that all the models are trained on different datasets. LCR and NCN are trained on Arxiv belonging to different years. Table \ref{tab:comparison} presents the results of the models across all diagnostic datasets, highlighting the top K results where K is 10 for both Recall and MRR. BM25 shows the best performance in terms of Recall and MRR for most datasets. Specifically, BM25 outperforms other models across all classes for diagnostic datasets categorized by fields, years, and citation counts. This comprehensive performance across different classes underscores the robustness of BM25. LCR achieves a better Recall value for the length dataset, while BM25 excels in MRR. In the location dataset, BM25 has the highest Recall when the reference is at the start of the context, whereas LCR performs better for references in the middle and end. However, BM25 consistently has a higher MRR across all classes in the location dataset. In the intent dataset, LCR demonstrates superior performance in Recall for all classes, whereas BM25 leads in MRR. For the low-resource dataset, BM25 dominates Recall and MRR across all classes. Additionally, BM25 achieves higher MRR across all classes and for BM25, except for the preceding Preposition and Others categories.

Due to the varying training datasets, different models perform differently on the classes of the diagnostic datasets. In the field datasets, NCN and LCR perform well in mathematics, while Galactica models perform well in computer science. On the other hand, BM25 performs the best in the chemistry domain. In the year's dataset, older years have had better performance in NCN and LCR, while Galactica models have better performance in recent years.

Similarly, BM25's performance peaks in the mid-level years. When it comes to citation count, a direct relationship is observed with the performance of NCN, which improves as the number of citation counts increases. Conversely, LCR demonstrates its strength for cited papers with a lower citation count. The performance of Galactica models is closely tied to the citation counts of the cited papers; in contrast, the performance of BM25 shows an inverse correlation with the citation count. NCN, LCR, and Galactica models show superior performance in the short-length context, while BM25 excels with longer contexts. LCR and Galactica models perform better for references positioned at the end of the context; on the other hand, NCN shows better performance for references in the middle. BM25 demonstrates the best overall performance when the reference is at the start of the context. In the intent dataset, NCN, LCR, and BM25 show the best performance for the intent with background class, while Galactica performs well for the methodology. Similar to the low-resource dataset, NCN shows better recall in philosophy, while MRR is better in business. For LCR and Galactica models, engineering demonstrates the best performance compared to other low-resource fields. BM25 shows better recall in art and MRR in geography. 
In the POS dataset, references with the following adjective POS perform better in NCN. For LCR, the preceding POS noun shows better performance. The Galactica models perform well with the preceding POS preposition, while in BM25, references with the preceding POS pronoun achieve higher performance.

As mentioned, the performance varies among different models, primarily due to differences in their training datasets and architectural designs. For instance, Galactica models exhibit lower performance in the location dataset when referencing the start and middle of sentences compared to the end. This discrepancy arises from Galactica's training on the Transformer model's decoder, which retains information about the preceding context before the masked token. Similarly, BM25 demonstrates superior overall performance compared to other models, primarily because it has the advantage of querying the entire S2AG dataset, which comprises millions of records.

\section{Limitations}
The size of the diagnostic dataset could be more significant, which was limited due to the time required to manually check the accuracy of the context. Additional datasets can be incorporated to evaluate more features of the context and global information. More published citation recommendation models could have been covered for the evaluations. Similarly, a standard framework, such as a library or script, could have been created to evaluate the models. Different evaluation metrics such as nDCG, F1 score, and perplexity could also be used. Similarly, models could also be compared based on parameters such as processing time, CPU utilization, memory requirement, and GPU utilization.

\section{Conclusion}

In conclusion, a comprehensive benchmark for evaluating and comparing citation recommendation models is established. Utilizing the S2ORC and S2AG datasets, thorough data cleaning and preprocessing have been performed, followed by the construction of diagnostic datasets based on various citation context features. These datasets have facilitated a robust assessment of model performance through relevant evaluation metrics. Additionally, the evaluation of selected citation recommendation models has enabled a standardized analysis and comparison of their performance. This study provides valuable insights into the effectiveness of various approaches within the field of citation recommendation, highlighting key strengths and areas for improvement in current methodologies.
% \section{Discussion, Challenges, and Limitations}

\bibliographystyle{ACM-Reference-Format}
\bibliography{sample-base}


\begin{thebibliography}{46}

%%% the Web address.

\ifx \showCODEN    \undefined \def \showCODEN     #1{\unskip}     \fi
\ifx \showDOI      \undefined \def \showDOI       #1{#1}\fi
\ifx \showISBNx    \undefined \def \showISBNx     #1{\unskip}     \fi
\ifx \showISBNxiii \undefined \def \showISBNxiii  #1{\unskip}     \fi
\ifx \showISSN     \undefined \def \showISSN      #1{\unskip}     \fi
\ifx \showLCCN     \undefined \def \showLCCN      #1{\unskip}     \fi
\ifx \shownote     \undefined \def \shownote      #1{#1}          \fi
\ifx \showarticletitle \undefined \def \showarticletitle #1{#1}   \fi
\ifx \showURL      \undefined \def \showURL       {\relax}        \fi
% invisible to TeX
\providecommand\bibfield[2]{#2}
\providecommand\bibinfo[2]{#2}
\providecommand\natexlab[1]{#1}
\providecommand\showeprint[2][]{arXiv:#2}

\bibitem[Ali et~al\mbox{.}(2022)]%
        {ali2022spr}
\bibfield{author}{\bibinfo{person}{Zafar Ali}, \bibinfo{person}{Guilin Qi}, \bibinfo{person}{Pavlos Kefalas}, \bibinfo{person}{Shah Khusro}, \bibinfo{person}{Inayat Khan}, {and} \bibinfo{person}{Khan Muhammad}.} \bibinfo{year}{2022}\natexlab{}.
\newblock \showarticletitle{SPR-SMN: Scientific paper recommendation employing SPECTER with memory network}.
\newblock \bibinfo{journal}{\emph{Scientometrics}} \bibinfo{volume}{127}, \bibinfo{number}{11} (\bibinfo{year}{2022}), \bibinfo{pages}{6763--6785}.
\newblock


\bibitem[Ali et~al\mbox{.}(2021)]%
        {ali2021overview}
\bibfield{author}{\bibinfo{person}{Zafar Ali}, \bibinfo{person}{Irfan Ullah}, \bibinfo{person}{Amin Khan}, \bibinfo{person}{Asim Ullah~Jan}, {and} \bibinfo{person}{Khan Muhammad}.} \bibinfo{year}{2021}\natexlab{}.
\newblock \showarticletitle{An overview and evaluation of citation recommendation models}.
\newblock \bibinfo{journal}{\emph{Scientometrics}} \bibinfo{volume}{126}, \bibinfo{number}{5} (\bibinfo{year}{2021}), \bibinfo{pages}{4083--4119}.
\newblock


\bibitem[Ammar et~al\mbox{.}(2018)]%
        {Ammar2018ConstructionOT}
\bibfield{author}{\bibinfo{person}{Bridger~Waleed Ammar}, \bibinfo{person}{Dirk Groeneveld}, \bibinfo{person}{Chandra Bhagavatula}, \bibinfo{person}{Iz Beltagy}, \bibinfo{person}{Miles Crawford}, \bibinfo{person}{Doug Downey}, \bibinfo{person}{Jason Dunkelberger}, \bibinfo{person}{Ahmed Elgohary}, \bibinfo{person}{Sergey Feldman}, \bibinfo{person}{Vu~A. Ha}, \bibinfo{person}{Rodney~Michael Kinney}, \bibinfo{person}{Sebastian Kohlmeier}, \bibinfo{person}{Kyle Lo}, \bibinfo{person}{Tyler~C. Murray}, \bibinfo{person}{Hsu-Han Ooi}, \bibinfo{person}{Matthew~E. Peters}, \bibinfo{person}{Joanna~L. Power}, \bibinfo{person}{Sam Skjonsberg}, \bibinfo{person}{Lucy~Lu Wang}, \bibinfo{person}{Christopher Wilhelm}, \bibinfo{person}{Zheng Yuan}, \bibinfo{person}{Madeleine van Zuylen}, {and} \bibinfo{person}{Oren Etzioni}.} \bibinfo{year}{2018}\natexlab{}.
\newblock \showarticletitle{Construction of the Literature Graph in Semantic Scholar}. In \bibinfo{booktitle}{\emph{North American Chapter of the Association for Computational Linguistics}}.
\newblock
\urldef\tempurl%
\url{https://api.semanticscholar.org/CorpusID:19170988}
\showURL{%
\tempurl}


\bibitem[Angrosh et~al\mbox{.}(2010)]%
        {angrosh2010context}
\bibfield{author}{\bibinfo{person}{MA Angrosh}, \bibinfo{person}{Stephen Cranefield}, {and} \bibinfo{person}{Nigel Stanger}.} \bibinfo{year}{2010}\natexlab{}.
\newblock \showarticletitle{Context identification of sentences in related work sections using a conditional random field: towards intelligent digital libraries}. In \bibinfo{booktitle}{\emph{Proceedings of the 10th annual joint conference on Digital libraries}}. \bibinfo{pages}{293--302}.
\newblock


\bibitem[Bethard and Jurafsky(2010)]%
        {bethard2010should}
\bibfield{author}{\bibinfo{person}{Steven Bethard} {and} \bibinfo{person}{Dan Jurafsky}.} \bibinfo{year}{2010}\natexlab{}.
\newblock \showarticletitle{Who should I cite: learning literature search models from citation behavior}. In \bibinfo{booktitle}{\emph{Proceedings of the 19th ACM international conference on Information and knowledge management}}. \bibinfo{pages}{609--618}.
\newblock


\bibitem[Bhagavatula et~al\mbox{.}(2018)]%
        {bhagavatula2018content}
\bibfield{author}{\bibinfo{person}{Chandra Bhagavatula}, \bibinfo{person}{Sergey Feldman}, \bibinfo{person}{Russell Power}, {and} \bibinfo{person}{Bridger~Waleed Ammar}.} \bibinfo{year}{2018}\natexlab{}.
\newblock \showarticletitle{Content-Based Citation Recommendation}.
\newblock \bibinfo{journal}{\emph{ArXiv}}  \bibinfo{volume}{abs/1802.08301} (\bibinfo{year}{2018}).
\newblock
\urldef\tempurl%
\url{https://api.semanticscholar.org/CorpusID:3536005}
\showURL{%
\tempurl}


\bibitem[Bird et~al\mbox{.}(2008)]%
        {bird2008acl}
\bibfield{author}{\bibinfo{person}{Steven Bird}, \bibinfo{person}{Robert Dale}, \bibinfo{person}{Bonnie~J Dorr}, \bibinfo{person}{Bryan~R Gibson}, \bibinfo{person}{Mark~Thomas Joseph}, \bibinfo{person}{Min-Yen Kan}, \bibinfo{person}{Dongwon Lee}, \bibinfo{person}{Brett Powley}, \bibinfo{person}{Dragomir~R Radev}, \bibinfo{person}{Yee~Fan Tan}, {et~al\mbox{.}}} \bibinfo{year}{2008}\natexlab{}.
\newblock \showarticletitle{The ACL Anthology Reference Corpus: A Reference Dataset for Bibliographic Research in Computational Linguistics.}. In \bibinfo{booktitle}{\emph{LREC}}.
\newblock


\bibitem[Boyack et~al\mbox{.}(2018)]%
        {boyack2018characterizing}
\bibfield{author}{\bibinfo{person}{Kevin~W Boyack}, \bibinfo{person}{Nees~Jan van Eck}, \bibinfo{person}{Giovanni Colavizza}, {and} \bibinfo{person}{Ludo Waltman}.} \bibinfo{year}{2018}\natexlab{}.
\newblock \showarticletitle{Characterizing in-text citations in scientific articles: A large-scale analysis}.
\newblock \bibinfo{journal}{\emph{Journal of Informetrics}} \bibinfo{volume}{12}, \bibinfo{number}{1} (\bibinfo{year}{2018}), \bibinfo{pages}{59--73}.
\newblock


\bibitem[Cai et~al\mbox{.}(2018)]%
        {cai2018three}
\bibfield{author}{\bibinfo{person}{Xiaoyan Cai}, \bibinfo{person}{Junwei Han}, \bibinfo{person}{Wenjie Li}, \bibinfo{person}{Renxian Zhang}, \bibinfo{person}{Shirui Pan}, {and} \bibinfo{person}{Libin Yang}.} \bibinfo{year}{2018}\natexlab{}.
\newblock \showarticletitle{A three-layered mutually reinforced model for personalized citation recommendation}.
\newblock \bibinfo{journal}{\emph{IEEE transactions on neural networks and learning systems}} \bibinfo{volume}{29}, \bibinfo{number}{12} (\bibinfo{year}{2018}), \bibinfo{pages}{6026--6037}.
\newblock


\bibitem[Cohan et~al\mbox{.}(2020)]%
        {cohan2020specter}
\bibfield{author}{\bibinfo{person}{Arman Cohan}, \bibinfo{person}{Sergey Feldman}, \bibinfo{person}{Iz Beltagy}, \bibinfo{person}{Doug Downey}, {and} \bibinfo{person}{Daniel~S. Weld}.} \bibinfo{year}{2020}\natexlab{}.
\newblock \showarticletitle{SPECTER: Document-level Representation Learning using Citation-informed Transformers}.
\newblock \bibinfo{journal}{\emph{ArXiv}}  \bibinfo{volume}{abs/2004.07180} (\bibinfo{year}{2020}).
\newblock
\urldef\tempurl%
\url{https://api.semanticscholar.org/CorpusID:215768677}
\showURL{%
\tempurl}


\bibitem[Dehghani et~al\mbox{.}(2021)]%
        {dehghani2021benchmark}
\bibfield{author}{\bibinfo{person}{Mostafa Dehghani}, \bibinfo{person}{Yi Tay}, \bibinfo{person}{Alexey~A. Gritsenko}, \bibinfo{person}{Zhe Zhao}, \bibinfo{person}{Neil Houlsby}, \bibinfo{person}{Fernando Diaz}, \bibinfo{person}{Donald Metzler}, {and} \bibinfo{person}{Oriol Vinyals}.} \bibinfo{year}{2021}\natexlab{}.
\newblock \showarticletitle{The Benchmark Lottery}.
\newblock \bibinfo{journal}{\emph{ArXiv}}  \bibinfo{volume}{abs/2107.07002} (\bibinfo{year}{2021}).
\newblock
\urldef\tempurl%
\url{https://api.semanticscholar.org/CorpusID:235810239}
\showURL{%
\tempurl}


\bibitem[Ebesu and Fang(2017)]%
        {ebesu2017neural}
\bibfield{author}{\bibinfo{person}{Travis Ebesu} {and} \bibinfo{person}{Yi Fang}.} \bibinfo{year}{2017}\natexlab{}.
\newblock \showarticletitle{Neural citation network for context-aware citation recommendation}. In \bibinfo{booktitle}{\emph{Proceedings of the 40th international ACM SIGIR conference on research and development in information retrieval}}. \bibinfo{pages}{1093--1096}.
\newblock


\bibitem[F{\"a}rber and Jatowt(2020)]%
        {farber2020citation}
\bibfield{author}{\bibinfo{person}{Michael F{\"a}rber} {and} \bibinfo{person}{Adam Jatowt}.} \bibinfo{year}{2020}\natexlab{}.
\newblock \showarticletitle{Citation recommendation: approaches and datasets}.
\newblock \bibinfo{journal}{\emph{International Journal on Digital Libraries}} \bibinfo{volume}{21}, \bibinfo{number}{4} (\bibinfo{year}{2020}), \bibinfo{pages}{375--405}.
\newblock


\bibitem[Gehrmann et~al\mbox{.}(2021)]%
        {gehrmann2021gem}
\bibfield{author}{\bibinfo{person}{Sebastian Gehrmann}, \bibinfo{person}{Tosin~P. Adewumi}, \bibinfo{person}{Karmanya Aggarwal}, \bibinfo{person}{Pawan~Sasanka Ammanamanchi}, \bibinfo{person}{Aremu Anuoluwapo}, \bibinfo{person}{Antoine Bosselut}, \bibinfo{person}{Khyathi~Raghavi Chandu}, \bibinfo{person}{Miruna Clinciu}, \bibinfo{person}{Dipanjan Das}, \bibinfo{person}{Kaustubh~D. Dhole}, \bibinfo{person}{Wanyu Du}, \bibinfo{person}{Esin Durmus}, \bibinfo{person}{Ondrej Dusek}, \bibinfo{person}{Chris~C. Emezue}, \bibinfo{person}{Varun Gangal}, \bibinfo{person}{Cristina Garbacea}, \bibinfo{person}{Tatsunori~B. Hashimoto}, \bibinfo{person}{Yufang Hou}, \bibinfo{person}{Yacine Jernite}, \bibinfo{person}{Harsh Jhamtani}, \bibinfo{person}{Yangfeng Ji}, \bibinfo{person}{Shailza Jolly}, \bibinfo{person}{Mihir Kale}, \bibinfo{person}{Dhruv Kumar}, \bibinfo{person}{Faisal Ladhak}, \bibinfo{person}{Aman Madaan}, \bibinfo{person}{Mounica Maddela}, \bibinfo{person}{Khyati Mahajan}, \bibinfo{person}{Saad Mahamood},
  \bibinfo{person}{Bodhisattwa~Prasad Majumder}, \bibinfo{person}{Pedro~Henrique Martins}, \bibinfo{person}{Angelina McMillan-Major}, \bibinfo{person}{Simon Mille}, \bibinfo{person}{Emiel van Miltenburg}, \bibinfo{person}{Moin Nadeem}, \bibinfo{person}{Shashi Narayan}, \bibinfo{person}{Vitaly Nikolaev}, \bibinfo{person}{Andre~Niyongabo Rubungo}, \bibinfo{person}{Salomey Osei}, \bibinfo{person}{Ankur~P. Parikh}, \bibinfo{person}{Laura Perez-Beltrachini}, \bibinfo{person}{Niranjan Rao}, \bibinfo{person}{Vikas Raunak}, \bibinfo{person}{Juan~Diego Rodriguez}, \bibinfo{person}{Sashank Santhanam}, \bibinfo{person}{Jo{\~a}o Sedoc}, \bibinfo{person}{Thibault Sellam}, \bibinfo{person}{Samira Shaikh}, \bibinfo{person}{Anastasia Shimorina}, \bibinfo{person}{Marco Antonio~Sobrevilla Cabezudo}, \bibinfo{person}{Hendrik Strobelt}, \bibinfo{person}{Nishant Subramani}, \bibinfo{person}{Wei Xu}, \bibinfo{person}{Diyi Yang}, \bibinfo{person}{Akhila Yerukola}, {and} \bibinfo{person}{Jiawei Zhou}.}
  \bibinfo{year}{2021}\natexlab{}.
\newblock \showarticletitle{The GEM Benchmark: Natural Language Generation, its Evaluation and Metrics}.
\newblock \bibinfo{journal}{\emph{ArXiv}}  \bibinfo{volume}{abs/2102.01672} (\bibinfo{year}{2021}).
\newblock
\urldef\tempurl%
\url{https://api.semanticscholar.org/CorpusID:231749968}
\showURL{%
\tempurl}


\bibitem[Giles et~al\mbox{.}(1998)]%
        {giles1998citeseer}
\bibfield{author}{\bibinfo{person}{C~Lee Giles}, \bibinfo{person}{Kurt~D Bollacker}, {and} \bibinfo{person}{Steve Lawrence}.} \bibinfo{year}{1998}\natexlab{}.
\newblock \showarticletitle{CiteSeer: An automatic citation indexing system}. In \bibinfo{booktitle}{\emph{Proceedings of the third ACM conference on Digital libraries}}. \bibinfo{pages}{89--98}.
\newblock


\bibitem[Gu et~al\mbox{.}(2022)]%
        {gu2022local}
\bibfield{author}{\bibinfo{person}{Nianlong Gu}, \bibinfo{person}{Yingqiang Gao}, {and} \bibinfo{person}{Richard~HR Hahnloser}.} \bibinfo{year}{2022}\natexlab{}.
\newblock \showarticletitle{Local citation recommendation with hierarchical-attention text encoder and SciBERT-based reranking}. In \bibinfo{booktitle}{\emph{Advances in Information Retrieval: 44th European Conference on IR Research, ECIR 2022, Stavanger, Norway, April 10--14, 2022, Proceedings, Part I}}. Springer, \bibinfo{pages}{274--288}.
\newblock


\bibitem[He et~al\mbox{.}(2010)]%
        {he2010context}
\bibfield{author}{\bibinfo{person}{Qi He}, \bibinfo{person}{Jian Pei}, \bibinfo{person}{Daniel Kifer}, \bibinfo{person}{Prasenjit Mitra}, {and} \bibinfo{person}{Lee Giles}.} \bibinfo{year}{2010}\natexlab{}.
\newblock \showarticletitle{Context-aware citation recommendation}. In \bibinfo{booktitle}{\emph{Proceedings of the 19th international conference on World wide web}}. \bibinfo{pages}{421--430}.
\newblock


\bibitem[Huang et~al\mbox{.}(2015)]%
        {huang2015neural}
\bibfield{author}{\bibinfo{person}{Wenyi Huang}, \bibinfo{person}{Zhaohui Wu}, \bibinfo{person}{Chen Liang}, \bibinfo{person}{Prasenjit Mitra}, {and} \bibinfo{person}{C Giles}.} \bibinfo{year}{2015}\natexlab{}.
\newblock \showarticletitle{A neural probabilistic model for context based citation recommendation}. In \bibinfo{booktitle}{\emph{Proceedings of the AAAI Conference on Artificial Intelligence}}, Vol.~\bibinfo{volume}{29}.
\newblock


\bibitem[Huang et~al\mbox{.}(2014)]%
        {huang2014refseer}
\bibfield{author}{\bibinfo{person}{Wenyi Huang}, \bibinfo{person}{Zhaohui Wu}, \bibinfo{person}{Prasenjit Mitra}, {and} \bibinfo{person}{C~Lee Giles}.} \bibinfo{year}{2014}\natexlab{}.
\newblock \showarticletitle{Refseer: A citation recommendation system}. In \bibinfo{booktitle}{\emph{IEEE/ACM joint conference on digital libraries}}. IEEE, \bibinfo{pages}{371--374}.
\newblock


\bibitem[Jeong et~al\mbox{.}(2020)]%
        {jeong2020context}
\bibfield{author}{\bibinfo{person}{Chanwoo Jeong}, \bibinfo{person}{Sion Jang}, \bibinfo{person}{Eunjeong Park}, {and} \bibinfo{person}{Sungchul Choi}.} \bibinfo{year}{2020}\natexlab{}.
\newblock \showarticletitle{A context-aware citation recommendation model with BERT and graph convolutional networks}.
\newblock \bibinfo{journal}{\emph{Scientometrics}}  \bibinfo{volume}{124} (\bibinfo{year}{2020}), \bibinfo{pages}{1907--1922}.
\newblock


\bibitem[Kinney et~al\mbox{.}(2023)]%
        {kinney2023semantic}
\bibfield{author}{\bibinfo{person}{Rodney~Michael Kinney}, \bibinfo{person}{Chloe Anastasiades}, \bibinfo{person}{Russell Authur}, \bibinfo{person}{Iz Beltagy}, \bibinfo{person}{Jonathan Bragg}, \bibinfo{person}{Alexandra Buraczynski}, \bibinfo{person}{Isabel Cachola}, \bibinfo{person}{Stefan Candra}, \bibinfo{person}{Yoganand Chandrasekhar}, \bibinfo{person}{Arman Cohan}, \bibinfo{person}{Miles Crawford}, \bibinfo{person}{Doug Downey}, \bibinfo{person}{Jason Dunkelberger}, \bibinfo{person}{Oren Etzioni}, \bibinfo{person}{Rob Evans}, \bibinfo{person}{Sergey Feldman}, \bibinfo{person}{Joseph Gorney}, \bibinfo{person}{David~W. Graham}, \bibinfo{person}{F.Q. Hu}, \bibinfo{person}{Regan Huff}, \bibinfo{person}{Daniel King}, \bibinfo{person}{Sebastian Kohlmeier}, \bibinfo{person}{Bailey Kuehl}, \bibinfo{person}{Michael Langan}, \bibinfo{person}{Daniel Lin}, \bibinfo{person}{Haokun Liu}, \bibinfo{person}{Kyle Lo}, \bibinfo{person}{Jaron Lochner}, \bibinfo{person}{Kelsey MacMillan}, \bibinfo{person}{Tyler~C.
  Murray}, \bibinfo{person}{Christopher Newell}, \bibinfo{person}{Smita~R Rao}, \bibinfo{person}{Shaurya Rohatgi}, \bibinfo{person}{Paul Sayre}, \bibinfo{person}{Zejiang Shen}, \bibinfo{person}{Amanpreet Singh}, \bibinfo{person}{Luca Soldaini}, \bibinfo{person}{Shivashankar Subramanian}, \bibinfo{person}{A. Tanaka}, \bibinfo{person}{Alex~D Wade}, \bibinfo{person}{Linda~M. Wagner}, \bibinfo{person}{Lucy~Lu Wang}, \bibinfo{person}{Christopher Wilhelm}, \bibinfo{person}{Caroline Wu}, \bibinfo{person}{Jiangjiang Yang}, \bibinfo{person}{Angele Zamarron}, \bibinfo{person}{Madeleine van Zuylen}, {and} \bibinfo{person}{Daniel~S. Weld}.} \bibinfo{year}{2023}\natexlab{}.
\newblock \showarticletitle{The Semantic Scholar Open Data Platform}.
\newblock \bibinfo{journal}{\emph{ArXiv}}  \bibinfo{volume}{abs/2301.10140} (\bibinfo{year}{2023}).
\newblock
\urldef\tempurl%
\url{https://api.semanticscholar.org/CorpusID:256194545}
\showURL{%
\tempurl}


\bibitem[Knoth and Zdrahal(2012)]%
        {knoth2012core}
\bibfield{author}{\bibinfo{person}{Petr Knoth} {and} \bibinfo{person}{Zdenek Zdrahal}.} \bibinfo{year}{2012}\natexlab{}.
\newblock \showarticletitle{CORE: three access levels to underpin open access}.
\newblock \bibinfo{journal}{\emph{D-Lib Magazine}} \bibinfo{volume}{18}, \bibinfo{number}{11/12} (\bibinfo{year}{2012}), \bibinfo{pages}{1--13}.
\newblock


\bibitem[Ko et~al\mbox{.}(2022)]%
        {ko2022survey}
\bibfield{author}{\bibinfo{person}{Hyeyoung Ko}, \bibinfo{person}{Suyeon Lee}, \bibinfo{person}{Yoonseo Park}, {and} \bibinfo{person}{Anna Choi}.} \bibinfo{year}{2022}\natexlab{}.
\newblock \showarticletitle{A survey of recommendation systems: recommendation models, techniques, and application fields}.
\newblock \bibinfo{journal}{\emph{Electronics}} \bibinfo{volume}{11}, \bibinfo{number}{1} (\bibinfo{year}{2022}), \bibinfo{pages}{141}.
\newblock


\bibitem[Kunnath et~al\mbox{.}(2020)]%
        {kunnath2020overview}
\bibfield{author}{\bibinfo{person}{Suchetha~N Kunnath}, \bibinfo{person}{David Pride}, \bibinfo{person}{Bikash Gyawali}, {and} \bibinfo{person}{Petr Knoth}.} \bibinfo{year}{2020}\natexlab{}.
\newblock \showarticletitle{Overview of the 2020 WOSP 3C citation context classification task}. In \bibinfo{booktitle}{\emph{Proceedings of the 8th International Workshop on Mining Scientific Publications}}. Association for Computational Linguistics, \bibinfo{pages}{75--83}.
\newblock


\bibitem[Lo et~al\mbox{.}(2020)]%
        {lo2019s2orc}
\bibfield{author}{\bibinfo{person}{Kyle Lo}, \bibinfo{person}{Lucy~Lu Wang}, \bibinfo{person}{Mark Neumann}, \bibinfo{person}{Rodney~Michael Kinney}, {and} \bibinfo{person}{Daniel~S. Weld}.} \bibinfo{year}{2020}\natexlab{}.
\newblock \showarticletitle{S2ORC: The Semantic Scholar Open Research Corpus}. In \bibinfo{booktitle}{\emph{Annual Meeting of the Association for Computational Linguistics}}.
\newblock
\urldef\tempurl%
\url{https://api.semanticscholar.org/CorpusID:215416146}
\showURL{%
\tempurl}


\bibitem[Lu et~al\mbox{.}(2023)]%
        {lu2023research}
\bibfield{author}{\bibinfo{person}{Yonghe Lu}, \bibinfo{person}{Meilu Yuan}, \bibinfo{person}{Jiaxin Liu}, {and} \bibinfo{person}{Minghong Chen}.} \bibinfo{year}{2023}\natexlab{}.
\newblock \showarticletitle{Research on semantic representation and citation recommendation of scientific papers with multiple semantics fusion}.
\newblock \bibinfo{journal}{\emph{Scientometrics}} \bibinfo{volume}{128}, \bibinfo{number}{2} (\bibinfo{year}{2023}), \bibinfo{pages}{1367--1393}.
\newblock


\bibitem[Medi{\'c} and {\v{S}}najder(2020)]%
        {medic2020improved}
\bibfield{author}{\bibinfo{person}{Zoran Medi{\'c}} {and} \bibinfo{person}{Jan {\v{S}}najder}.} \bibinfo{year}{2020}\natexlab{}.
\newblock \showarticletitle{Improved local citation recommendation based on context enhanced with global information}. In \bibinfo{booktitle}{\emph{Proceedings of the First Workshop on Scholarly Document Processing}}. \bibinfo{pages}{97--103}.
\newblock


\bibitem[Nogueira et~al\mbox{.}(2020)]%
        {nogueira2020navigation}
\bibfield{author}{\bibinfo{person}{Rodrigo Nogueira}, \bibinfo{person}{Zhiying Jiang}, \bibinfo{person}{Kyunghyun Cho}, {and} \bibinfo{person}{Jimmy Lin}.} \bibinfo{year}{2020}\natexlab{}.
\newblock \showarticletitle{Navigation-based candidate expansion and pretrained language models for citation recommendation}.
\newblock \bibinfo{journal}{\emph{Scientometrics}} \bibinfo{volume}{125}, \bibinfo{number}{3} (\bibinfo{year}{2020}), \bibinfo{pages}{3001--3016}.
\newblock


\bibitem[Ren et~al\mbox{.}(2014)]%
        {ren2014cluscite}
\bibfield{author}{\bibinfo{person}{Xiang Ren}, \bibinfo{person}{Jialu Liu}, \bibinfo{person}{Xiao Yu}, \bibinfo{person}{Urvashi Khandelwal}, \bibinfo{person}{Quanquan Gu}, \bibinfo{person}{Lidan Wang}, {and} \bibinfo{person}{Jiawei Han}.} \bibinfo{year}{2014}\natexlab{}.
\newblock \showarticletitle{Cluscite: Effective citation recommendation by information network-based clustering}. In \bibinfo{booktitle}{\emph{Proceedings of the 20th ACM SIGKDD international conference on Knowledge discovery and data mining}}. \bibinfo{pages}{821--830}.
\newblock


\bibitem[Ritchie et~al\mbox{.}(2008)]%
        {ritchie2008comparing}
\bibfield{author}{\bibinfo{person}{Anna Ritchie}, \bibinfo{person}{Stephen Robertson}, {and} \bibinfo{person}{Simone Teufel}.} \bibinfo{year}{2008}\natexlab{}.
\newblock \showarticletitle{Comparing citation contexts for information retrieval}. In \bibinfo{booktitle}{\emph{Proceedings of the 17th ACM conference on Information and knowledge management}}. \bibinfo{pages}{213--222}.
\newblock


\bibitem[Saier and F{\"a}rber(2019)]%
        {saier2019bibliometric}
\bibfield{author}{\bibinfo{person}{Tarek Saier} {and} \bibinfo{person}{Michael F{\"a}rber}.} \bibinfo{year}{2019}\natexlab{}.
\newblock \showarticletitle{Bibliometric-Enhanced arXiv: A Data Set for Paper-Based and Citation-Based Tasks.}. In \bibinfo{booktitle}{\emph{BIR@ ECIR}}. \bibinfo{pages}{14--26}.
\newblock


\bibitem[Saier and F{\"a}rber(2020)]%
        {saier2020unarxive}
\bibfield{author}{\bibinfo{person}{Tarek Saier} {and} \bibinfo{person}{Michael F{\"a}rber}.} \bibinfo{year}{2020}\natexlab{}.
\newblock \showarticletitle{unarXive: a large scholarly data set with publications’ full-text, annotated in-text citations, and links to metadata}.
\newblock \bibinfo{journal}{\emph{Scientometrics}} \bibinfo{volume}{125}, \bibinfo{number}{3} (\bibinfo{year}{2020}), \bibinfo{pages}{3085--3108}.
\newblock


\bibitem[Tang et~al\mbox{.}(2008)]%
        {tang2008arnetminer}
\bibfield{author}{\bibinfo{person}{Jie Tang}, \bibinfo{person}{Jing Zhang}, \bibinfo{person}{Limin Yao}, \bibinfo{person}{Juanzi Li}, \bibinfo{person}{Li Zhang}, {and} \bibinfo{person}{Zhong Su}.} \bibinfo{year}{2008}\natexlab{}.
\newblock \showarticletitle{Arnetminer: extraction and mining of academic social networks}. In \bibinfo{booktitle}{\emph{Proceedings of the 14th ACM SIGKDD international conference on Knowledge discovery and data mining}}. \bibinfo{pages}{990--998}.
\newblock


\bibitem[Ta{\c{s}}k{\i}n and Al(2018)]%
        {tacskin2018content}
\bibfield{author}{\bibinfo{person}{Zehra Ta{\c{s}}k{\i}n} {and} \bibinfo{person}{Umut Al}.} \bibinfo{year}{2018}\natexlab{}.
\newblock \showarticletitle{A content-based citation analysis study based on text categorization}.
\newblock \bibinfo{journal}{\emph{Scientometrics}} \bibinfo{volume}{114}, \bibinfo{number}{1} (\bibinfo{year}{2018}), \bibinfo{pages}{335--357}.
\newblock


\bibitem[Taylor et~al\mbox{.}(2022)]%
        {taylor2022galactica}
\bibfield{author}{\bibinfo{person}{Ross Taylor}, \bibinfo{person}{Marcin Kardas}, \bibinfo{person}{Guillem Cucurull}, \bibinfo{person}{Thomas Scialom}, \bibinfo{person}{Anthony~S. Hartshorn}, \bibinfo{person}{Elvis Saravia}, \bibinfo{person}{Andrew Poulton}, \bibinfo{person}{Viktor Kerkez}, {and} \bibinfo{person}{Robert Stojnic}.} \bibinfo{year}{2022}\natexlab{}.
\newblock \showarticletitle{Galactica: A Large Language Model for Science}.
\newblock \bibinfo{journal}{\emph{ArXiv}}  \bibinfo{volume}{abs/2211.09085} (\bibinfo{year}{2022}).
\newblock
\urldef\tempurl%
\url{https://api.semanticscholar.org/CorpusID:253553203}
\showURL{%
\tempurl}


\bibitem[Te et~al\mbox{.}(2022)]%
        {te2022citation}
\bibfield{author}{\bibinfo{person}{Sonita Te}, \bibinfo{person}{Amira Barhoumi}, \bibinfo{person}{Martin Lentschat}, \bibinfo{person}{Fr{\'e}d{\'e}rique Bordignon}, \bibinfo{person}{Cyril Labb{\'e}}, {and} \bibinfo{person}{Fran{\c{c}}ois Portet}.} \bibinfo{year}{2022}\natexlab{}.
\newblock \showarticletitle{Citation Context Classification: Critical vs Non-critical}. In \bibinfo{booktitle}{\emph{Proceedings of the Third Workshop on Scholarly Document Processing}}. \bibinfo{pages}{49--53}.
\newblock


\bibitem[Thierry et~al\mbox{.}(2023)]%
        {thierry2023rar}
\bibfield{author}{\bibinfo{person}{Nimbeshaho Thierry}, \bibinfo{person}{Bing-Kun Bao}, {and} \bibinfo{person}{Zafar Ali}.} \bibinfo{year}{2023}\natexlab{}.
\newblock \showarticletitle{RAR-SB: research article recommendation using SciBERT with BiGRU}.
\newblock \bibinfo{journal}{\emph{Scientometrics}} \bibinfo{volume}{128}, \bibinfo{number}{12} (\bibinfo{year}{2023}), \bibinfo{pages}{6427--6448}.
\newblock


\bibitem[Trost(1986)]%
        {trost1986statistically}
\bibfield{author}{\bibinfo{person}{Jan~E Trost}.} \bibinfo{year}{1986}\natexlab{}.
\newblock \showarticletitle{Statistically nonrepresentative stratified sampling: A sampling technique for qualitative studies}.
\newblock \bibinfo{journal}{\emph{Qualitative sociology}} \bibinfo{volume}{9}, \bibinfo{number}{1} (\bibinfo{year}{1986}), \bibinfo{pages}{54--57}.
\newblock


\bibitem[Wang et~al\mbox{.}(2019)]%
        {wang2019superglue}
\bibfield{author}{\bibinfo{person}{Alex Wang}, \bibinfo{person}{Yada Pruksachatkun}, \bibinfo{person}{Nikita Nangia}, \bibinfo{person}{Amanpreet Singh}, \bibinfo{person}{Julian Michael}, \bibinfo{person}{Felix Hill}, \bibinfo{person}{Omer Levy}, {and} \bibinfo{person}{Samuel Bowman}.} \bibinfo{year}{2019}\natexlab{}.
\newblock \showarticletitle{Superglue: A stickier benchmark for general-purpose language understanding systems}.
\newblock \bibinfo{journal}{\emph{Advances in neural information processing systems}}  \bibinfo{volume}{32} (\bibinfo{year}{2019}).
\newblock


\bibitem[Wang et~al\mbox{.}(2018)]%
        {wang2018glue}
\bibfield{author}{\bibinfo{person}{Alex Wang}, \bibinfo{person}{Amanpreet Singh}, \bibinfo{person}{Julian Michael}, \bibinfo{person}{Felix Hill}, \bibinfo{person}{Omer Levy}, {and} \bibinfo{person}{Samuel~R. Bowman}.} \bibinfo{year}{2018}\natexlab{}.
\newblock \showarticletitle{GLUE: A Multi-Task Benchmark and Analysis Platform for Natural Language Understanding}. In \bibinfo{booktitle}{\emph{BlackboxNLP@EMNLP}}.
\newblock
\urldef\tempurl%
\url{https://api.semanticscholar.org/CorpusID:5034059}
\showURL{%
\tempurl}


\bibitem[Wang et~al\mbox{.}(2020a)]%
        {wang2020content}
\bibfield{author}{\bibinfo{person}{Leipeng Wang}, \bibinfo{person}{Yuan Rao}, \bibinfo{person}{Qinyu Bian}, {and} \bibinfo{person}{Shuo Wang}.} \bibinfo{year}{2020}\natexlab{a}.
\newblock \showarticletitle{Content-based hybrid deep neural network citation recommendation method}. In \bibinfo{booktitle}{\emph{International conference of pioneering computer scientists, engineers and educators}}. Springer, \bibinfo{pages}{3--20}.
\newblock


\bibitem[Wang et~al\mbox{.}(2020b)]%
        {wang2020important}
\bibfield{author}{\bibinfo{person}{Mingyang Wang}, \bibinfo{person}{Jiaqi Zhang}, \bibinfo{person}{Shijia Jiao}, \bibinfo{person}{Xiangrong Zhang}, \bibinfo{person}{Na Zhu}, {and} \bibinfo{person}{Guangsheng Chen}.} \bibinfo{year}{2020}\natexlab{b}.
\newblock \showarticletitle{Important citation identification by exploiting the syntactic and contextual information of citations}.
\newblock \bibinfo{journal}{\emph{Scientometrics}}  \bibinfo{volume}{125} (\bibinfo{year}{2020}), \bibinfo{pages}{2109--2129}.
\newblock


\bibitem[Wright and Augenstein(2021)]%
        {wright2021citeworth}
\bibfield{author}{\bibinfo{person}{Dustin Wright} {and} \bibinfo{person}{Isabelle Augenstein}.} \bibinfo{year}{2021}\natexlab{}.
\newblock \showarticletitle{CiteWorth: Cite-Worthiness Detection for Improved Scientific Document Understanding}. In \bibinfo{booktitle}{\emph{Findings}}.
\newblock
\urldef\tempurl%
\url{https://api.semanticscholar.org/CorpusID:235166702}
\showURL{%
\tempurl}


\bibitem[Yang et~al\mbox{.}(2018)]%
        {yang2018lstm}
\bibfield{author}{\bibinfo{person}{Libin Yang}, \bibinfo{person}{Yu Zheng}, \bibinfo{person}{Xiaoyan Cai}, \bibinfo{person}{Hang Dai}, \bibinfo{person}{Dejun Mu}, \bibinfo{person}{Lantian Guo}, {and} \bibinfo{person}{Tao Dai}.} \bibinfo{year}{2018}\natexlab{}.
\newblock \showarticletitle{A LSTM based model for personalized context-aware citation recommendation}.
\newblock \bibinfo{journal}{\emph{IEEE access}}  \bibinfo{volume}{6} (\bibinfo{year}{2018}), \bibinfo{pages}{59618--59627}.
\newblock


\bibitem[Yin and Li(2017)]%
        {yin2017personalized}
\bibfield{author}{\bibinfo{person}{Jun Yin} {and} \bibinfo{person}{Xiaoming Li}.} \bibinfo{year}{2017}\natexlab{}.
\newblock \showarticletitle{Personalized citation recommendation via convolutional neural networks}. In \bibinfo{booktitle}{\emph{Web and Big Data: First International Joint Conference, APWeb-WAIM 2017, Beijing, China, July 7--9, 2017, Proceedings, Part II 1}}. Springer, \bibinfo{pages}{285--293}.
\newblock


\bibitem[Zhu et~al\mbox{.}(2015)]%
        {zhu2015measuring}
\bibfield{author}{\bibinfo{person}{Xiaodan Zhu}, \bibinfo{person}{Peter Turney}, \bibinfo{person}{Daniel Lemire}, {and} \bibinfo{person}{Andr{\'e} Vellino}.} \bibinfo{year}{2015}\natexlab{}.
\newblock \showarticletitle{Measuring academic influence: Not all citations are equal}.
\newblock \bibinfo{journal}{\emph{Journal of the Association for Information Science and Technology}} \bibinfo{volume}{66}, \bibinfo{number}{2} (\bibinfo{year}{2015}), \bibinfo{pages}{408--427}.
\newblock


\end{thebibliography}

\end{document}